\documentclass[journal]{IEEEtran}
\usepackage{cite}
\usepackage{amsmath,amssymb,amsfonts}
\usepackage{algorithmic}
\usepackage{graphicx}
\usepackage[dvipsnames]{xcolor}
\usepackage{url}
\usepackage{soul}
\usepackage{textcomp}
\def\BibTeX{{\rm B\kern-.05em{\sc i\kern-.025em b}\kern-.08em
    T\kern-.1667em\lower.7ex\hbox{E}\kern-.125emX}}
    
\makeatother
\graphicspath{{img/}}
\DeclareGraphicsExtensions{.pdf}
\usepackage[binary-units=true]{siunitx}
\DeclareSIUnit{\sample}{S}
\DeclareSIUnit{\msps}{\mega\sample\per\s}
\DeclareSIUnit{\gsps}{\giga\sample\per\s}
\DeclareSIUnit{\gbps}{\giga b\per\s}
\DeclareSIUnit{\baud}{Bd}

\begin{document}

\title{Design and Experimental Verification of a Novel
  Error-Backpropagation-Based Background Calibration for Time
  Interleaved ADC in Digital Communication Receivers}
\author{\IEEEauthorblockN{{Fredy Solis}\IEEEauthorrefmark{1},
  \IEEEmembership{Student Member, IEEE}, {Benjam\'in T. Reyes\IEEEauthorrefmark{1}, Dami\'an A. Morero\IEEEauthorrefmark{2}, and Mario R. Hueda}\IEEEauthorrefmark{2}} 
  \IEEEauthorblockA {\\
    \IEEEauthorrefmark{1} Fundaci\'on Fulgor - Romagosa 518 - C\'ordoba (5000) - Argentina \\
    \IEEEauthorrefmark{2} Laboratorio de Comunicaciones Digitales - Universidad Nacional de  C\'ordoba\\
    Av. V\'elez Sarsfield 1611 - C\'ordoba (X5016GCA) - Argentina\\
    Email: fsolis@fundacionfulgor.org.ar}}

%

\markboth{IEEE TRANSACTIONS ON CIRCUITS AND SYSTEMS--I: REGULAR
  PAPERS,~Vol.~, No.~, }%
{Solis \MakeLowercase{\textit{et al.}}: Design and Experimental Verification of a Novel
  Error-Backpropagation-Based Background Calibration for Time
  Interleaved ADC in Digital Communication Receivers}

\maketitle

\begin{abstract}
  A novel background calibration technique for Time-Interleaved
  Analog-to-Digital Converters (TI-ADCs) is presented in this paper.
  This technique is applicable to equalized digital communication
  receivers. As shown by Tsai et al.~\cite{tsai_correction_2009} and
  Luna et al. ~\cite{luna_compensation_2006}, in a digital receiver it
  is possible to treat the TI-ADC errors as part of the communication
  channel and take advantage of the adaptive equalizer to compensate
  them. Therefore calibration becomes an integral part of the channel
  equalization. No special purpose analog or digital calibration
  blocks or algorithms are required. However, there is a large class
  of receivers where the equalization technique cannot be directly
  applied because other signal processing blocks are located between
  the TI-ADC and the equalizer. The technique presented here
  generalizes earlier works to this class of receivers. The error
  backpropagation algorithm, traditionally used in machine learning,
  is applied to the error computed at the receiver slicer and used to
  adapt an auxiliary equalizer adjacent to the TI-ADC, called the
  Compensation Equalizer (CE).  Simulations using a dual polarization
  optical coherent receiver model demonstrate accurate and robust
  mismatch compensation across different application
  scenarios. Several Quadrature Amplitude Modulation (QAM) schemes are
  tested in simulations and experimentally.  Measurements on an
  emulation platform which includes an 8 bit, 4 GS/s TI-ADC prototype
  chip fabricated in 130nm CMOS technology, show an almost ideal
  mitigation of the impact of the mismatches on the receiver
  performance when 64-QAM and 256-QAM schemes are tested.  An absolute
  improvement in the TI-ADC performance of \SI{\sim15}{\dB} in both
  SNDR and SFDR is measured.
\end{abstract}




\section{Introduction}
\label{sec:intro}
\IEEEPARstart{T}{his} paper proposes a novel background calibration
technique for Time-Interleaved Analog-to-Digital Converters
(TI-ADCs)used in equalized digital communication receivers. It
generalizes a previously proposed technique
\cite{tsai_correction_2009,luna_compensation_2006,
  agazzi_90_2008}. Current and emerging digital receivers for ultra
high-speed communication systems~\cite{morero_design_2016,
  faruk_digital_2017, crivelli_architecture_2014, palermo_cmos_2016,
  ethernet_alliance_2020_2020, brandolini_5_2015,xu_5-bit_2016}
require large bandwidth, high sampling rate Analog-to-Digital
Converters (ADCs).

The TI-ADC~\cite{laperle_advances_2014,kull_cmos_2016} has been the
technique predominantly used to meet the demanding sampling rate and
bandwidth requirements of high-speed transceivers.

The performance of TI-ADCs is affected by mismatches among the
interleaves~\cite{kurosawa_explicit_2001,vogel_impact_2005}. Mismatches
of sampling time, gain, bandwidth, as well as DC offset, are the most
common impairments. Many calibration techniques have been proposed in
the literature. Please see \cite{wei_8_2014, song_10-b_2017,
  mafi_digital_2017, guo_16-gss_2020, lee_1_2014,
  saleem_adaptive_2010, matsuno_all-digital_2013,
  harpe_oversampled_2014, ali_12-b_2020, reyes_joint_2012,
  reyes_design_2017, le_duc_fully_2017, guo_5_2020,
  salib_high-precision_2019,
  murmann_digitally_2013,harpe_digitally_2015} and references therein
for a thorough review and discussion.

Calibration techniques for general purpose TI-ADCs in general require
dedicated calibration blocks and algorithms. On the other hand,
several authors \cite{ luna_compensation_2006, tsai_correction_2009,
  agazzi_90_2008} have shown that in the special case of an equalized
digital receiver, it is possible to treat the TI-ADC errors as
integral part of the communication channel and take advantage of the
already existing adaptive equalizer to compensate them. Therefore
calibration becomes an integral part of the channel equalization. No
special purpose analog or digital calibration blocks or algorithms are
required.  Equalizer-based compensation can compensate static as well
as frequency-dependent errors such as bandwidth limitations and
bandwidth or frequency response mismatches among the interleaves of
the ADC. Because the equalizer is adaptive, it also compensates
time-dependent effects such as those caused by temperature and voltage
variations, or by aging. Therefore, equalization becomes the TI-ADC
compensation technique of choice in digital communication receivers.

\begin{figure}
  \centering
  \includegraphics[width=1.\columnwidth]{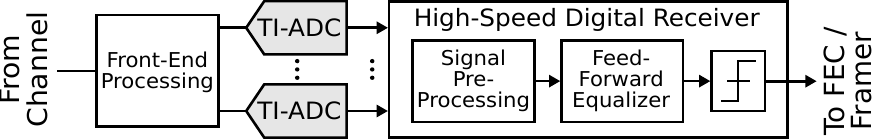}
  \caption{\label{f:typ_rx_adc}Typical high-speed digital
    receiver. This scheme can be found in wireline, wireless, and
    optical fiber communications. The receiver may use more than one
    ADC.}
\end{figure}

Tsai et al~\cite{tsai_correction_2009} provide a thorough description
of the equalization technique and its advantages. However, there is a
large class of receivers where this technique cannot be directly
applied because other signal processing blocks are located between the
TI-ADC and the equalizer. A block diagram of a typical receiver for
high-speed digital communications is shown in Fig.~\ref{f:typ_rx_adc}.

An effective compensation of the TI-ADCs errors has been achieved in
the referenced works \cite{ luna_compensation_2006,
  tsai_correction_2009, agazzi_90_2008} because the main receiver
equalizer, or {Feedforward Equalizer} (FFE) is immediately located
after the TI-ADC (in other words, the Signal Pre-Processing block of
Fig.~\ref{f:typ_rx_adc} is not present). Hence, the FFE can access and
directly compensate the impairments of the different interleaves.
Also, the slicer error carries information about the impairments of
the individual interleaves and therefore the FFE adaptation algorithm
can drive its coefficients to a solution that jointly compensates the
channel and the TI-ADC impairments.  Unfortunately, the application of
Tsai's technique to most types of receivers (e.g., for coherent
optical communications) has been limited by the presence of signal
pre-processing blocks (e.g., Timing Recovery (TR), Carrier Recovery
(CR), or {Bulk Chromatic Dispersion Equalizer} (BCD)).  These blocks
cause signal components associated with different interleaves of the
TI-ADC to be combined in a way that makes the use of the FFE
unsuitable to compensate them.

The main contribution of this work is a new background
technique~\cite{solis_background_2020} that overcomes the
aforementioned limitations, and is especially well suited for complex
digital receivers.  The basic idea consists in the use of an
auxiliary, low complexity adaptive equalizer, called the
\emph{Compensation Equalizer} (CE), to compensate the mismatches of
the TI-ADC. Slicer error components associated with
different interleaves are
also combined by the signal pre-processing blocks. Thus, the
slicer error is not directly applicable to adapt the CE.

In this work we propose to adapt the CE using a post processed version
of the error at the slicer of the receiver. The post processing is
based on the \emph{backpropagation
  algorithm}~\cite{rumelhart_learning_1986}, widely used in machine
learning applications~\cite{goodfellow_deep_2016}.  Its main
characteristic is that, in a multi-stage processing chain where
several cascaded blocks have adaptive parameters, it is able to
determine the contribution to the error generated by each one of these
blocks and their associated parameters for all the stages.
Backpropagation is used in combination with the Stochastic Gradient
Algorithm (SGD) to adjust the coefficients of the CE in order to
minimize the slicer Mean Squared Error (MSE).  The use of the CE in
combination with the backpropagation algorithm results in robust, fast
converging background calibration.  As we shall show, this proposal is
not limited to the compensation of individual TI-ADCs (which is the
case for most calibration techniques), but it extends itself to the
entire receiver Analog Front End (AFE), enabling the compensation of
impairments such as time skew, quadrature, and amplitude errors
between the in-phase and the quadrature components of the signal in a
receiver based on Phase Modulation (PM) or QAM.

Because ultrafast adaptation is usually not needed, the
backpropagation algorithm can be implemented in a highly subsampled
hardware block which does not require parallel processing. Therefore,
the implementation complexity of the proposed technique is low, as
will be discussed in detail.  Although the technique presented here is
general and can be used in digital receivers for different
applications, the primary example in this paper is a receiver for
coherent optical communications. State of the art coherent optical
receivers operate at symbol rates around 96 Giga-Baud (GBd) and
require ADC sampling rates close to \SI{150}{\gsps} and bandwidths of
about \SI{50}{\GHz}. In the near future symbol rates will increase to
\SIrange{128}{150}{\giga\baud} or higher, requiring bandwidths in the
range of \SIrange{65}{75}{\GHz} and sampling rates in the
\SIrange{200}{250}{\gsps} range. High-order Quadrature Amplitude
Modulation (QAM) schemes (e.g., 64-QAM, 256-QAM and higher) will be
deployed to increase spectral
efficiency~\cite{raybon_high_2015}. High-order modulation schemes
increase the resolution and overall performance requirements on the
ADC.

The benefits of the proposed technique are experimentally verified
using 64-QAM and 256-QAM schemes. The core of the experimental setup
is an \SI{8}{bit}, \SI{4}{\gsps} TI-ADC test
chip~\cite{solis_4gss_2021}.

The rest of this paper is organized as follows.
Section~\ref{sec:System_Model} presents a discrete time model of the
TI-ADC system in a Dual-Polarization (DP) optical coherent
receiver. The error backpropagation based adaptive CE is introduced in
Section~\ref{sec:EBP}. Simulation results are presented and discussed
in Section~\ref{s:sim_res}.  The experimental evaluation is performed
in Section~\ref{s:experimental}, and conclusions are drawn in
Section~\ref{s:conclusion}.

\section{System Model of High-Speed Digital Receivers based on TI-ADC}
\label{sec:System_Model}

\begin{figure}
  \centering
  \includegraphics[width=1.\columnwidth]{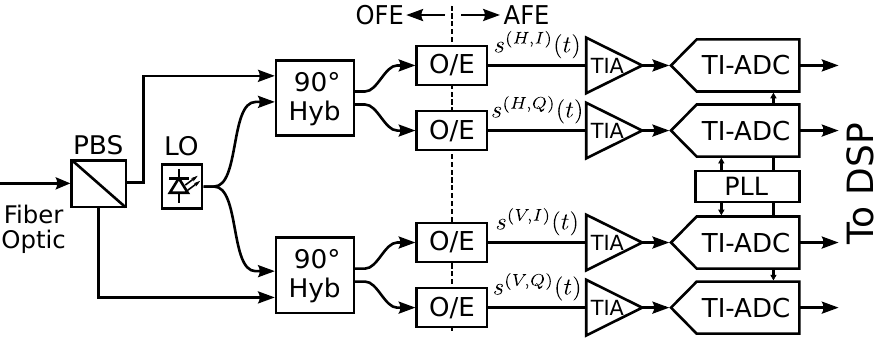}
  \caption{\label{f:ofe} Optical/analog front-end for a TI-ADC-based
    coherent optical receiver. The optical signal is split into four
    electrical lanes that are converted by a TI-ADC. {PBS}:
    Polarization Beam Splitter; {LO}: Local Oscillator;
    {$90^o$ Hyb}: $90^o$ hybrid coupler.}
\end{figure}

Communication channels of interest in this work include, among others:
\textit{i)}~wireline, \textit{ii)}~wireless, or \textit{iii)}~optical.
The primary example of application of the backpropagation-based
compensation technique considered in the next sections is a DP 
coherent optical receiver
~\cite{morero_design_2016,faruk_digital_2017,
  crivelli_architecture_2014}. However, it can be used in any
high-speed digital receiver with minor modifications.  A block diagram
of the Optical Front End (OFE) and the AFE for a DP coherent receiver
is shown in Fig.~\ref{f:ofe}.  The optical input signal is decomposed
by the OFE into four signals, the in-phase and quadrature ($I/Q$)
components of the horizontal and vertical ($H/V$) polarizations.
Photodetectors are used to convert the optical signals to
photocurrents which are amplified by {Trans-Impedance Amplifiers}
(TIAs).  Then, the AFE acquires the electrical signals and translates
them to the digital domain.  Digital receivers with a certain degree
of oversampling (e.g., $T_s=\frac{T}{2}$ where $T_s$ and $T$ are the
sampling and symbol periods, respectively) are used to compensate the
dispersion experienced in optical links~\cite{crivelli_adaptive_2004}.
Next, we formulate the model of the optical channel, including a
TI-ADC system affected by mismatches, used in the remainder of this
paper.

Let
$a_k^{({\mathcal P})}=a_k^{({\mathcal P},{ I})}+ja_k^{({\mathcal
    P},{Q})}$ be the transmitted QAM symbol in polarization
${\mathcal P}\in\{H,V\}$ at time instant $k$.  The Chromatic
Dispersion (CD) and Polarization-Mode Dispersion (PMD) effects of an
optical fiber link can be modeled as a $2 \times 2$ {Multiple-Input
  Multiple-Output} (MIMO) complex-valued
channel~\cite{crivelli_adaptive_2004} encompassing four complex
filters with impulse responses ${\overline h}_{m,n}(t)$ where
$m,n=1,2$.  For a comprehensive description of the effects of the
optical channel, please see~\cite{agrawal_fiberoptic_2010}.

Then, the noise-free electrical signals provided by the optical
demodulator in the receiver can be expressed
as~\cite{crivelli_adaptive_2004}
\begin{align}
  \label{eq:eq0H}
  s^{(H)}(t)&=s^{(H,I)}(t)+js^{(H,Q)}(t)\\
  \nonumber
            &=e^{j\omega_0 t}\left[\sum_k a_k^{(H)}{\overline h}_{1,1}(t-kT)+ a_k^{(V)}{\overline h}_{1,2}(t-kT)\right],
  \label{eq:eq0V}
\end{align}
\begin{align}
  s^{(V)}(t)&=s^{(V,I)}(t)+js^{(V,Q)}(t)\\
  \nonumber
            &=e^{j\omega_0t}\left[\sum_k a_k^{(H)}{\overline h}_{2,1}(t-kT)+ a_k^{(V)}{\overline h}_{2,2}(t-kT)\right],
\end{align}
where $\omega_0$ is the optical carrier frequency offset (or frequency
difference between the transmitter and the local oscillator) and $1/T$
is the symbol rate.

\subsection{AFE and TI-ADC Discrete-Time Model}
\label{ss:afe_tiadc_model}
\begin{figure}
  \centering
  \includegraphics[width=1.\columnwidth]{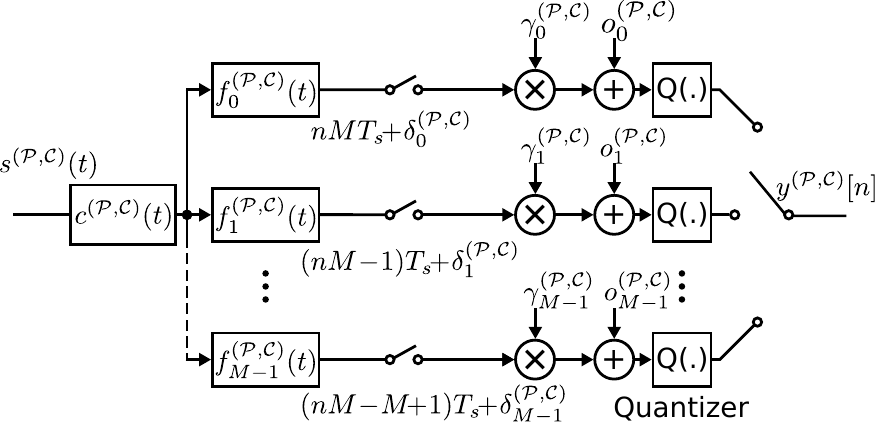}
  \caption{\label{f:fig1} Analog front-end model for polarization
    $\mathcal P \in\{H,V\}$ and component $\mathcal C \in\{I,Q\}$ in a
    TI-ADC-based DP coherent optical receiver.}
\end{figure}

A discrete-time model for the AFE and the TI-ADC system of
Fig. \ref{f:ofe} with their impairments is introduced in this section.
A simplified representation of the analog path for one component
${\mathcal C}\in\{I,Q\}$ in a given polarization
${\mathcal P\in\{H,V\}}$ is shown in Fig.~\ref{f:fig1}.  The response
of the electrical interconnections between the optical demodulator and
the TIA, the TIA response itself, and any other components in the
signal path up to a TI-ADC system is represented with a filter with
impulse response $c^{({\mathcal P},{\mathcal C})}(t)$.  Time delay or
\emph{skew} between components $I$ and $Q$ of a given polarization
$\mathcal P$ is caused by mismatches between $c^{({\mathcal P},I)}(t)$
and $c^{({\mathcal P},Q)}(t)$, and degrades the receiver performance.
As we shall show, the proposed background calibration algorithm is
able to compensate not only the mismatches of the TI-ADC, but also the
I/Q skew, the quadrature and amplitude errors, and other impairments
among the signal paths.

Blocks $f^{({\mathcal P},{\mathcal C})}_{m}(t)$ ($m=0,\cdots, M-1$)
model the independent responses of the $M$ Track and Hold (T\&H)
circuits in an $M$-channel TI-ADC system\footnote{Typically the
  frequency responses of the T\&H can be assumed as a first-order
  low-pass filter~\cite{sin_statistical_2008}. Such response arises
  from the combination of the on-resistance of a CMOS switch with the
  sampling capacitor of the ADC, or the input capacitance of an analog
  buffer.}.  Each one of the $M$ interleaved channels is sampled every
$M/f_s=MT_s$ seconds with a proper sampling phase.  Sampling time
errors and the DC offsets are represented with
$\delta_m^{({\mathcal P},{\mathcal C})}$ and
$o^{({\mathcal P},{\mathcal C})}_m$, respectively.  Every path
gain/attenuation is modeled by
\begin{equation}
  \label{eq:gain_error}
  \gamma^{({\mathcal P},{\mathcal C})}_m=1+\Delta_{\gamma^{({\mathcal
        P},{\mathcal C})}_m},
\end{equation}
where $\Delta_{\gamma^{({\mathcal P},{\mathcal C})}_m}$ is the gain error.


The quantizer is modeled as additive white noise
with uniform distribution since the resolution of the ADC is
considered sufficiently high.  Also, at high-frequency (i.e.,
$1/T_s$), the DC offsets $o^{({\mathcal P},{\mathcal C})}_m$ generate
an $M$-periodic signal denoted as
${\tilde o}^{({\mathcal P},{\mathcal C})}[n]$ such that
${\tilde o}^{({\mathcal P},{\mathcal C})}[n]={\tilde o}^{({\mathcal
    P},{\mathcal C})}[n+M]$ with
\begin{equation}
  \label{eq:tilde)}
  {\tilde o}^{({\mathcal P},{\mathcal C})}[m]={o}^{({\mathcal
      P},{\mathcal C})}_m,\quad m=0,\cdots,M-1.
\end{equation}

The digitized high-frequency samples can be written as (see Appendix
A)
\begin{align}
  \nonumber
  y^{({\mathcal P},{\mathcal C})}[n]=&\sum_{l} {\tilde h}^{({\mathcal P},{\mathcal C})}_n[l] s^{({\mathcal P},{\mathcal C})}[n-l]+{\tilde o}^{({\mathcal P},{\mathcal C})}[n]+\\
  \label{eq:eq1b}
              &q^{({\mathcal P},{\mathcal C})}[n].
\end{align}
where ${\tilde h}^{({\mathcal P},{\mathcal C})}_n[l]$ is the impulse
response of a time-varying filter, which is an $M$-periodic sequence
such
${\tilde h}^{({\mathcal P},{\mathcal C})}_n[l]={\tilde h}^{({\mathcal
    P},{\mathcal C})}_{n+M}[l]$ defined by \eqref{eq:eq5}, and
$q^{({\mathcal P},{\mathcal C})}[n]$ is the quantization noise.

\subsection{Compensation of AFE Mismatch and TI-ADC Impairments}




Errors and mismatches of the TI-ADC can be compensated by using
digital finite impulse response (FIR) filters applied to each
interleaved branch. In the case of a communication receiver, the
digitized signal could be applied to a time-varying equalizer
immediately following the TI-ADC (see~\cite{tsai_correction_2009} for
more details).
The practical implementation of this periodically time-varying
equalizer is briefly addressed in Section III-A, and in more detail
in~\cite{agazzi_90_2008}.

Similarly to what was done in previous
works~\cite{luna_compensation_2006,agazzi_90_2008,tsai_correction_2009,
  saleem_adaptive_2010}, in the backpropagation-based architecture
introduced in this paper we propose to adaptively compensate the
TI-ADC mismatch, after the mitigation of the offset, using a filter
with an $M$-periodic time-varying impulse response:
\begin{equation}
  \label{eq:eq6}
  x^{({\mathcal P},{\mathcal C})}[n]=\sum_{l=0}^{L_g-1} {\tilde g}^{({\mathcal P},{\mathcal C})}_n[l]
  {w}^{({\mathcal P},{\mathcal C})}[n-l],
\end{equation}
where ${\tilde g}^{({\mathcal P},{\mathcal C})}_n[l]$ is the
$M$-periodic time-varying impulse response of the compensation filter
(i.e.,
${\tilde g}^{({\mathcal P},{\mathcal C})}_n[l]={\tilde g}^{({\mathcal
    P},{\mathcal C})}_{n+M}[l]$), $L_g$ is the number of taps of the
compensation filters, and $w^{({\mathcal P},{\mathcal C})}[n]$ is the
DC offset-free signal given by
\begin{equation}
  \label{eq:w}
  w^{({\mathcal P},{\mathcal C})}[n]=y^{({\mathcal P},{\mathcal
      C})}[n]-{\hat {\tilde o}}^{({\mathcal P},{\mathcal C})}[n],
\end{equation}
with ${\hat {\tilde o}}^{({\mathcal P},{\mathcal C})}[n]$ being the
$M$-periodic offset sequence estimation.  The combination of the
offset compensation blocks and the compensation filters
${\tilde g}^{({\mathcal P},{\mathcal C})}_n[l]$ constitutes the
CE (see Fig.~\ref{f:f5}).

The adaptation algorithm of the CE as proposed in
\cite{tsai_correction_2009} or \cite{luna_compensation_2006} cannot be
implemented in coherent optical communication receivers. This is
because of the presence of several signal pre-processing blocks placed
between the CE and the slicers, such as the Bulk Chromatic Dispersion
Equalizer (BCD) or the MIMO FFE that compensates PMD
\cite{morero_design_2016}. Thus, since the slicer errors are not
available at the outputs of the CE, a proper strategy has to be
defined to adapt the CE response.

On the other hand, it is worth to highlight that mismatches between the
$I$ and $Q$ signal paths cannot be mitigated using adaptive
compensation techniques based on a reference ADC, such
as~\cite{saleem_adaptive_2010}.  In the next section, we apply the
\emph{backpropagation} technique to adapt the CE coefficients.

\section{Error-Backpropagation-Based Compensation of AFE and TI-ADC
  Impairments in Digital Receivers}
\label{sec:EBP}

\begin{figure}[t]
  \centering
  \includegraphics[width=1.\columnwidth]{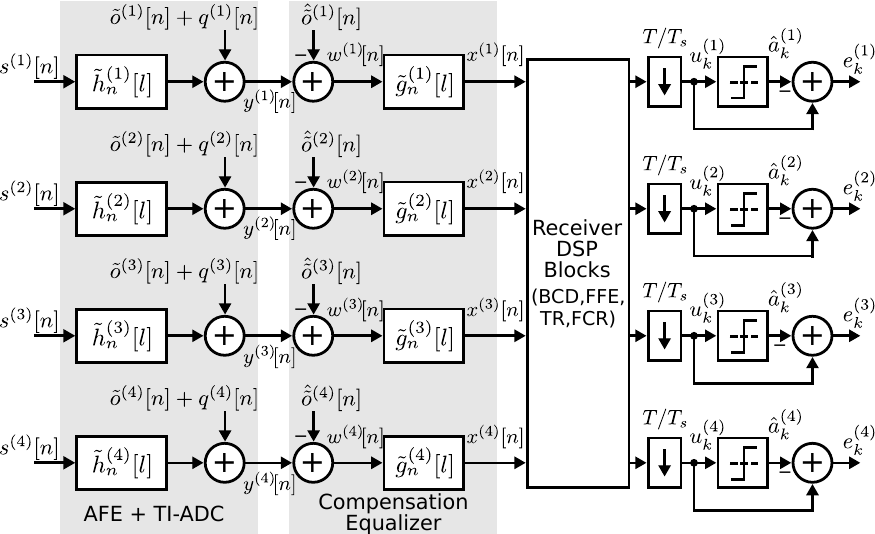}
  \caption{\label{f:f5} Block diagram of a DP optical
    coherent receiver with the CE for mitigating the effects
    of both AFE mismatches and TI-ADC impairments.}
\end{figure}

A block diagram of the AFE+TI-ADC in a DP optical coherent receiver
with the adaptive compensation equalizer, including four instances of
the real filter as defined by~\eqref{eq:eq6}, is depicted in
Fig.~\ref{f:f5}.  Please notice that the notation of previous figures
has been modified for simplicity. In Fig.~\ref{f:f5} an integer index
between 1 and 4 is used to differentiate a certain component in a
given polarization. Signals $s^{(H,I)}[n]$ and $s^{(H,Q)}[n]$ are
represented by $s^{(1)}[n]$ and $s^{(2)}[n]$, respectively.
Similarly, $s^{(3)}[n]$ and $s^{(4)}[n]$ represent $s^{(V,I)}[n]$ and
$s^{(V,Q)}[n]$, respectively.

The {Digital Signal Processing} (DSP) block of Fig.~\ref{f:f5},
performs the main receiver functions, operating with samples every
$T_s$ seconds.  In summary, some of the most important DSP algorithms
used in optical coherent receivers are the BCD, the MIMO FFE, TR from
the received symbols, the {Fine Carrier Recovery} (FCR) to compensate
the carrier phase and frequency offset\footnote{Although the receiver
  DSP for wireline and wireless may include other algorithms, the
  technique presented here can be applied to them with minor
  modifications.}.  Readers interested in more details on optical
coherent receivers can see~\cite{morero_design_2016,
  faruk_digital_2017, fludger_digital_2014} and references therein.

\subsection{Parallel Implementation of the Compensation Equalizer}
Before explaining the adaptation of the CE
coefficients ${\tilde g}^{(i)}_{n}[l]$, it is important to highlight
that no additional complexity is added to implement the CE with
independent responses multiplexed in time when they are implemented as
a parallel architecture. Let ${g}^{(i)}_m[l]$ with $i=1,\cdots,4$ be
the filter impulse response ${\tilde g}^{(i)}_{n}[l]$ in one period
defined as
\begin{equation}
  \label{eq:gPC}
  {g}^{(i)}_m[l]={\tilde g}^{(i)}_{m+n_0}[l], \quad m=0,\cdots,M-1,
\end{equation}
where $l=0,\cdots,L_g-1$ and $n_0$ is an arbitrary time index multiple
of $M$. Thus, notice that the application of the CE in coherent
receivers (see Fig. \ref{f:f5}) comprises 4 sets of real valued Finite
Impulse Response (FIR) ${g}_m^{(i)}[l]$ with $i=1,2,3,4$,
$m=0,\cdots,M-1$, and $l=0,\cdots, L_g-1$.

In high speed optical communication applications, the use of parallel
implementations is mandatory.  Typically, a parallelism factor on the
order of 128 or higher is adopted.  Furthermore, given the number of
interleaves of the TI-ADC $M$, the parallelism factor $P$ can be
selected to be a multiple of $M$, i.e., $P=q\times M$ with $q$ an
integer.  In this way, the different time multiplexed taps are located
in fixed positions of the parallel implementation, and we do not incur
significant additional complexity when compared to a filter with just
one set of coefficients (see~\cite{luna_compensation_2006} for more
details). The complexity of the resulting filter is similar to that of
the I/Q-skew compensation filter already present in current coherent
receivers~\cite{morero_design_2016}.  Moreover, the typical skew
correction filter can be replaced by the CE without adding significant
penalties in area or power since our proposal is also able to correct
time skew.

\subsection{All Digital Compensation Architecture}
\label{ss:dig_comp}

The filter coefficients of the impulse response
in~\eqref{eq:gPC} are adapted using the slicer error at the output of
the receiver DSP block.  We denote $e_k^{(j)}$ as the \emph{slicer
  error}, defined as
\begin{equation}
  \label{eq:ePC}
  e_k^{(j)}=u_k^{(j)}-{\hat a}_k^{(j)},\quad j=1,\cdots,4,
\end{equation}
where $u_k^{(j)}$ and ${\hat a}_k^{(j)}$ are the $k$-th slicer input
and output, respectively (see Fig.~\ref{f:f5}).  ${\hat a}_k^{(j)}$ is
also called the detected symbol.  Since the slicer operates at $1/T$
sampling rate, a subsampling of $T/T_s$ is needed after the receiver
DSP block.  Then, the total squared error at the slicer at time
instant $k$ is defined as
\begin{equation}
  \label{eq:eT}
  {\mathcal E}_k=\sum_{j=1}^4|e_k^{(j)}|^2.
\end{equation}

Let $E\{{\mathcal E}_k\}$ be the MSE at the slicer with $E\{.\}$
denoting the expectation operator.  In this work we iteratively adapt
the real coefficients of the CE defined by \eqref{eq:gPC} by using the
{Least Mean Squares} (LMS) algorithm, in order to minimize the MSE at
the slicer:
\begin{equation}
  \label{eq:lmsg}
  {\mathbf g}^{(i)}_{m,p+1}={\mathbf g}^{(i)}_{m,p}- \beta
  \nabla_{{\mathbf g}^{(i)}_{m,p}} E\{{\mathcal
    E}_k\},
\end{equation}
where $i=1,\cdots,4$; $m=0,\cdots,M-1$; $p$ denotes the number of
iteration, ${\mathbf g}^{(i)}_{m,p}$ is the $L_g$-dimensional
coefficient vector at the $p$-th iteration given by
\begin{equation}
  \label{eq:vgPC}
  {\mathbf g}^{(i)}_{m,p}=\left[{g}^{(i)}_{m,p}[0], {g}^{(i)}_{m,p}[1], \cdots, {g}^{(i)}_{m,p}[L_g-1]  \right]^T,
\end{equation}
where $\beta$ is the adaptation step, and
$\nabla_{{\mathbf g}^{(i)}_{m,p}} E\{{\mathcal E}_k\}$ is the gradient
of the MSE with respect to the vector ${\mathbf g}^{(i)}_{m,p}$.

We highlight that the key obstacle of the previous analysis is the
computation of the MSE gradient since ${\mathcal E}_k$ is not the
error at the output of the CE block.  To address this problem, we
propose the use of the \emph{backpropagation algorithm}, extensively
used in \emph{machine learning}
applications~\cite{rumelhart_learning_1986, goodfellow_deep_2016}.  By
applying this algorithm to the slicer errors, we are now able to
generate the error samples needed to adapt the coefficients of the
filters, as expressed in~\eqref{eq:lmsg}.  Consequently, the gradient
$\nabla_{{\mathbf g}^{(i)}_{m,p}} E\{{\mathcal E}_k\}$ can be
estimated as usual in the traditional LMS algorithm, using these
\textit{backpropagated} errors.

\subsection{Error Backpropagation}
\label{s:bp_formulation}
\begin{figure}
  \centering \includegraphics[width=1.\columnwidth]{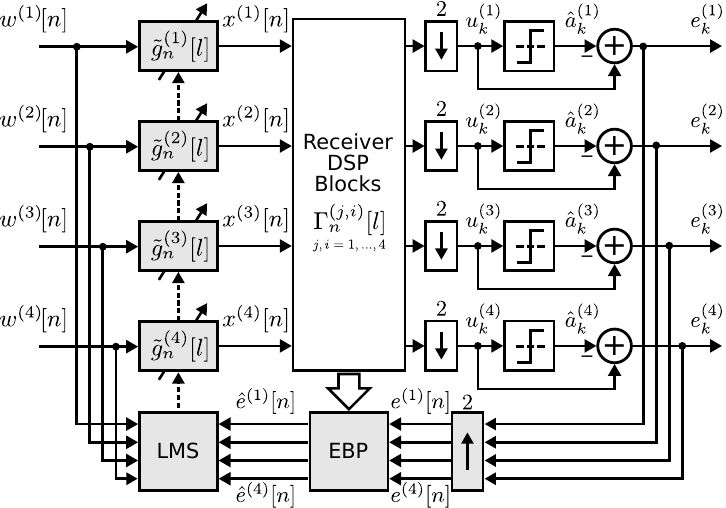}
  \caption{\label{f:f6} Block diagram of the proposed Error
    Backpropagation (EBP) based adaptation architecture for AFE+TI-ADC
    impairments compensation in a DP optical coherent receiver with
    $T/T_s=2$.}
\end{figure}

Without loss of generality, we consider that the receiver DSP block
can be modeled as a real time-varying $4 \times 4$ MIMO $T/2$
fractionally spaced equalizer (i.e., $T_s=T/2$), which is able to
compensate CD and PMD among other optical fiber channel effects.
Then, we can write the downsampled output of the $T/2$ receiver DSP
block (see Fig~\ref{f:f6}) as
\begin{equation}
  \label{eq:u1}
  u^{(j)}_k=\sum_{i=1}^4\sum_{l=0}^{L_{\Gamma}-1} {\Gamma}^{(j,i)}_{2k}[l]{x}^{(i)}[2k-l],\quad j=1,\cdots,4,
\end{equation}
where ${\Gamma}_n^{(j,i)}[l]$ is the time-varying impulse response of
the filter with input $i$ and output $j$, $L_{\Gamma}$ is the number
of coefficients of the filter, whereas ${x}^{(i)}[l]$ is the signal at
the DSP block input $i$ given by~\eqref{eq:eq6}, i.e.,
\begin{equation}
  \label{eq:eq6b}
  x^{(i)}[n]=\sum_{l'=0}^{L_g-1} {g}^{(i)}_{\lfloor n\rfloor_M}[l']
  {w}^{(i)}[n-l'],\quad i=1,\cdots,4,
\end{equation}
where ${g}^{(i)}_m$ is the impulse response defined by~\eqref{eq:gPC},
$\lfloor .\rfloor_M$ denotes the modulo $M$ operation, and
${w}^{(i)}[n]$ is the DC compensated signal given by~\eqref{eq:w}.

The gradient of the MSE,
$\nabla_{{\mathbf g}^{(i)}_{m,p}} E\{{\mathcal E}_k\}$ can be replaced
by a noisy estimation $\nabla_{{\mathbf g}^{(i)}_{m}} {\mathcal E}_k$,
as usual with the SGD based adaptation.  As we show in
Appendix~\ref{app:backprop}, an \emph{instantaneous} gradient of the
squared error~\eqref{eq:eT} can be expressed as
\begin{equation}
  \label{eq:grad}
  \nabla_{{\mathbf g}^{(i)}_{m}} {\mathcal E}_k=\alpha {\hat e}^{(i)}[m+kM]{\mathbf w}^{(i)}[m+kM],
\end{equation}
where $\alpha$ is a certain constant, ${\mathbf w}[n]$ is a vector
with $L_g$ input samples of the CE, i.e.,
\begin{equation}
  \label{eq:vecw}
  {\mathbf w}^{(i)}[n]=\left[{w}^{(i)}[n],{w}^{(i)}[n-1],\cdots,{w}^{(i)}[n-L_g+1]  \right]^T,
\end{equation}
where the \emph{backpropagated error}, ${\hat e}^{(i)}[n]$ is
expressed as
\begin{equation}
  \label{eq:bpe}
  {\hat e}^{(i)}[n]=\sum_{j=1}^4\sum_{l=0}^{L_{\Gamma}-1}\Gamma^{(j,i)}_{n+l}[l] e^{(j)}[n+l],
\end{equation}
with $e^{(j)}[n]$ being the \emph{oversampled} slicer error generated
from the slicer error at the \emph{baud-rate} $e_k^{(j)}$
in~\eqref{eq:ePC} as
\begin{equation}
  \label{eq:oe}
  e^{(j)}[n] = 
  \begin{cases} 
    e_{n/2}^{(j)}              & \mbox{if } n= 0,\pm 2,\pm 4,\cdots   \\
    0 & \mbox{otherwise}
  \end{cases}.
\end{equation}
Then, we can derive an all-digital compensation scheme using an
adaptive CE with coefficients updated as
\begin{equation}
  \label{eq:lmsg2}
  {\mathbf g}^{(i)}_{m,p+1}={\mathbf g}^{(i)}_{m,p}- \mu \nabla_{{\mathbf g}^{(i)}_{m,p}} {\mathcal E}_k,
\end{equation}
where $\mu=\alpha \beta$ is the adaptation step-size. Moreover, it is
possible to estimate the DC offsets in the input samples, using the
backpropagated error defined in~\eqref{eq:bpe}, as follows
\begin{equation}
  {\hat { {o}}}^{(i)}_{m,p+1}={\hat { {o}}}^{(i)}_{m,p}-\mu_o{\hat
    e}^{(i)}[n+m], \quad m=0,\cdots, M-1,
  \label{eq:tildeo}
\end{equation}
where {${\hat { {o}}}^{(i)}_{m,p}$} is the DC offset sequence estimation
in one period (see~\eqref{eq:w}) at the $p$-th iteration, and $\mu_o$
is the step-size of the DC offset estimator.

In order to avoid possible instability due to competition between the
CE and any adaptive DSP blocks in $\Gamma_n^{j,i}[l]$ (e.g., the
FFE), an adaptation constraint must be included.  This can be achieved
by limiting one of the $4M$ sets of the CE coefficients to only be a
time delay line.  For example, ${g}^{(0)}_{0}[l]=\delta_{l,l_d}$ where
$l=0,\cdots,L_g-1$ and $l_d=\frac{L_g+1}{2}$ ($L_g$ is assumed odd).

The coefficient updates given by \eqref{eq:lmsg2} and
\eqref{eq:tildeo} do not need to operate at full rate, because channel
impairments change slowly over time.  Then, subsampling can be
applied.  In this way, the implementation complexity can be
significantly reduced.  Further complexity reduction is enabled by:
\textit{i}) strobing the algorithms once they have converged, and/or
\textit{ii}) implementing them in firmware in an embedded processor,
typically available in coherent optical transceivers.  


\subsection{Mixed-Signal Calibration Architecture}
\label{s:mixed}

\begin{figure}
  \centering
  \includegraphics[width=.9\columnwidth]{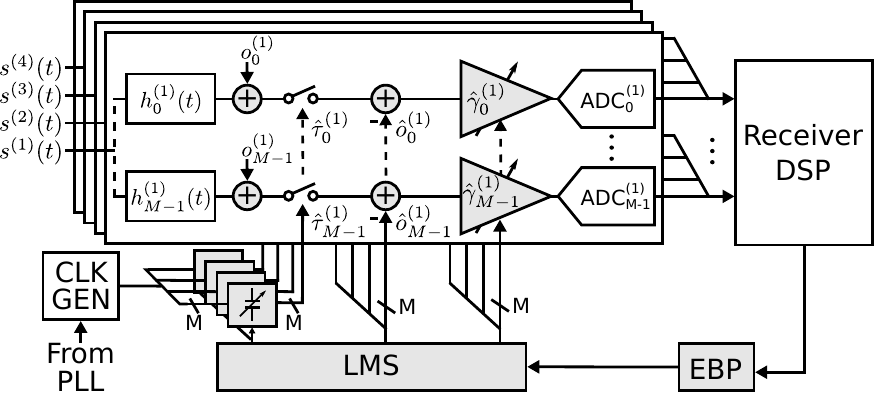}
  \caption{\label{f:ms_cal_sch}Block diagram of the mixed-signal
    calibration variant. The calibration with analog elements enables
    power consumption reduction.  The EBP is the same as the
    all-digital variant. {For simplicity the iteration index
    is omitted in the calibration parameters.}}
\end{figure}

The Error Backpropagation (EBP) algorithm just described also enables
a mixed-signal calibration technique.  A block diagram of this
calibration is depicted in Fig.~\ref{f:ms_cal_sch}.  With this
variant, sampling phase, gain, and DC offsets are adjusted prior to
the ADC\footnote{Notice that, in contrast to the full-digital
  compensation variant, the described mixed-signal solution requires
  additional considerations in order to compensate some effects such
  as bandwidth mismatches.} by using the gradient of the
backpropagated slicer error.  With this calibration approach the DC
offsets are compensated using~\eqref{eq:tildeo}, similar to the full
digital variant.  The gain coefficient is updated using
\begin{equation}
  \hat \gamma^{(i)}_{m,p+1}=\hat \gamma^{(i)}_{m,p}- \mu_{\gamma}
  {\hat e}^{(i)}[m+kM]w^{(i)}[m+kM],\quad \forall k,
\end{equation}
where $m=0,\cdots,M-1$ and $i=1,\cdots,4$.  Finally, the sampling
phase can be calibrated using the {MMSE timing recovery
  algorithm}~\cite{lee_digital_2004}, since the backpropagated slicer
error is available at the ADC outputs, i.e.,
\begin{align}
  \hat \tau^{(i)}_{m,p+1}=&\hat \tau^{(i)}_{m,p}-\mu_{\tau}{\hat e}^{(i)}[m+kM]\times\\
  \nonumber
                          &\left(w^{(i)}[m+kM+1]-w^{(i)}[m+kM-1]\right),\ 
                            \forall k
\end{align}
with $m=0,\cdots,M-1$. The calibration algorithm has the advantage of
tuning analog elements already present in most implementations of the
TI-ADC~\cite{reyes_design_2017,
  reyes_energy-efficient_2019,kull_24-72-gs/s_2018}.  For example, the
clock sampling phase is adjusted with variable delay lines, gain and
offset can be corrected in the comparator or with Programmable Gain
Amplifiers (PGA), if required.


\section{Simulation Results}
\label{s:sim_res}
\begin{table}[t]
\centering
\caption{Parameters Used in Simulations (UDRVD: Uniformly Distributed
  Random Variable. VFS: Full-Scale Voltage).}
\label{t:sim_parameters}
  \begin{tabular}{l|c}
    \hline
    {\textbf{Parameter}}                              & \textbf{Value}
    \\ \hline 
    Modulation                                        & 64-QAM         \\ 
    Symbol Rate ($f_B=1/T$)                           & 96 GBd         \\ 
    Receiver Oversampling Factor ($T/T_s$)            & 2            \\ 
    Fiber Length                                      & 100 km \\ 
    Differential Group Delay (DGD)                    & 10 ps          \\ 
    Second Order Pol. Mode Disp. (SOPMD) & 1000 ps$^2$    \\ 
    Speed of Rotation of the Pol. at the Tx   & 2 kHz           \\ 
    Speed of Rotation of the Pol. at the Rx   & 10 kHz          \\ 
    TI-ADC Resolution                                 & 8 bit          \\ 
    TI-ADC Sampling Rate (all interleaves)            & 192 GS/s        \\ 
    Number of Interleaves of TI-ADC ($M$)             & 16             \\ 
    Number of Taps of CE ($L_g$)  & 7\\
    Roll-off Factor  & 0.10\\
    Nominal BW of Analog Paths ($B_0$) (see \eqref{eq:B}) & 53 GHz\\ 
    Gain Errors (see \eqref{eq:gain_error}) - UDRV & $\Delta_{\gamma^{(i)}_m}\in [\pm 0.15]$ \\
    Sampling Phase Errors - UDRV  &  $\delta_m^{(i)} \in [\pm 0.075]T$\\
    Bandwidth Mismatches (see \eqref{eq:B}) - UDRV &  $\Delta_{B_{m}^{(i)}} \in [\pm 0.075]B_0$\\
    I/Q Time Skew  - UDRV &  $\tau_H,\tau_V\in [\pm 0.075]T$\\
    DC Offsets - UDRV  &  $o_m^{(i)} \in [\pm 0.025]$VFS\\
    \\ \hline
  \end{tabular}
\end{table}
\begin{figure}[t]
  \centering
  \includegraphics[width=\columnwidth]{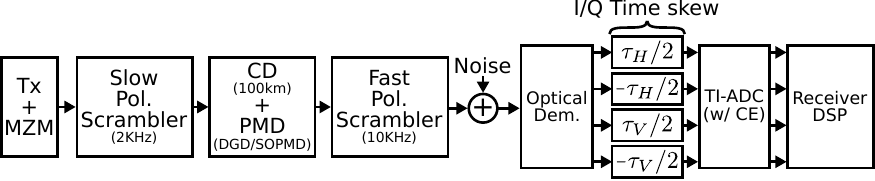}
  \caption{\label{f:simulation_setup}Block diagram of the system model
    used in simulations.}
\end{figure}

In this section the proposed backpropagation based mismatch
compensation technique is tested using simulations.  The simulation
setup is shown in Fig.~\ref{f:simulation_setup}.  The simulated
parameters are summarized in Table~\ref{t:sim_parameters}.  TI-ADC
mismatches are modeled as Uniformly Distributed Random Variables
(UDRV).  The electrical analog path responses~\eqref{eq:eq2} are
modeled by first-order low-pass filters with 3dB-bandwidth defined by
\begin{equation}
  \label{eq:B}
  B^{(i)}_m=B_0+\Delta_{B^{(i)}_m},\quad i=1,2,3,4;\quad 
  m=0,\cdots,M-1, 
\end{equation}
where $B_0$ is the nominal BW and $\Delta_{B^{(i)}_m}$ is the BW
mismatch.  Sampling phase errors and I/Q time skew are modeled by
Lagrange interpolation filters.  The I/Q time skew of each
polarization is evenly distributed between its corresponding
components (see Fig.~\ref{f:simulation_setup}).  Errors of the TI-ADC
are modeled as detailed in Section~\ref{ss:afe_tiadc_model}. In
particular, time skews among the interleaves are modeled using
Lagrange interpolation filters (not to be confused with those used to
model the I/Q skews).  We consider a DP optical coherent system with a
64-QAM modulation scheme, and a symbol rate of
$1/T=\SI{96}{\giga\baud}$.  Raised cosine filters with roll-off factor
$0.10$ for transmit pulse shaping are simulated (i.e., the nominal BW
of the channel filters is
$B_0=1.1\times \frac{\SI{96}{\GHz}}{2}\approx \SI{53}{\GHz}$).  The
Optical Signal-to-Noise Ratio (OSNR) is set to that required to
achieve a Bit-Error-Rate (BER) of $\sim 1\times 10^{-3}$
(see~\cite{freude_quality_2012,chan_optical_2010} for the definition
of OSNR).  The oversampling factor in the DSP blocks is $T/T_s=2$.
The fiber length is \SI{100}{\km} with \SI{10}{\ps} of Differential
Group Delay (DGD) and \SI{1000}{ps^2} of Second-Order PMD (SOPMD).
Rotations of the State of Polarization (SOP) of \SI{2}{\kHz} and
\SI{10}{\kHz} are included at the transmitter and receiver,
respectively.  Please see~\cite{agrawal_fiberoptic_2010} for a
comprehensive description of the aforementioned optical channel
parameters.  TI-ADCs with 8-bit resolution,
\SI{192}{\giga\sample\per\s} sampling rate, and $M=16$ are simulated.
The number of taps of the digital compensation filters is $L_g=7$.

\subsection{Montecarlo Simulations of the Adaptive CE}
\label{sec:montecarlo}
\begin{figure}
\centering
\includegraphics[width=.8\columnwidth]{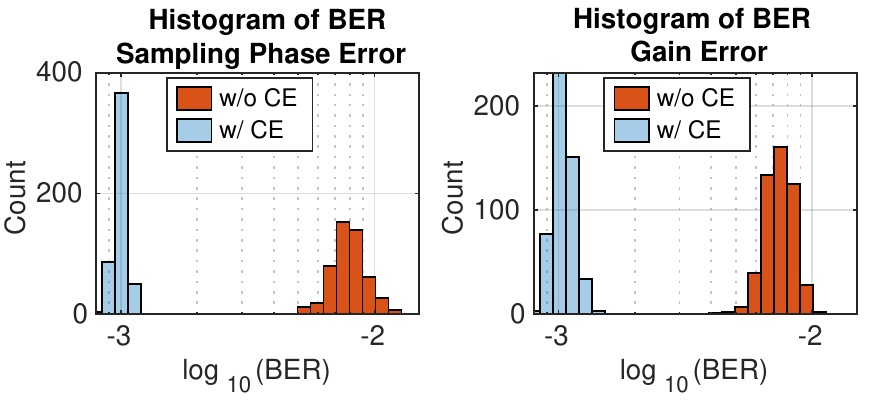}
\caption{\label{f:hist1}Histogram of the BER for 500 random cases with
  and without the CE for a reference BER of $\sim 1\times 10^{-3}$. Left:
  sampling phase errors (only). Right: gain errors (only).  See
  simulation parameters in Table~\ref{t:sim_parameters}.}
\end{figure}

\begin{figure}
\centering
\includegraphics[width=.8\columnwidth]{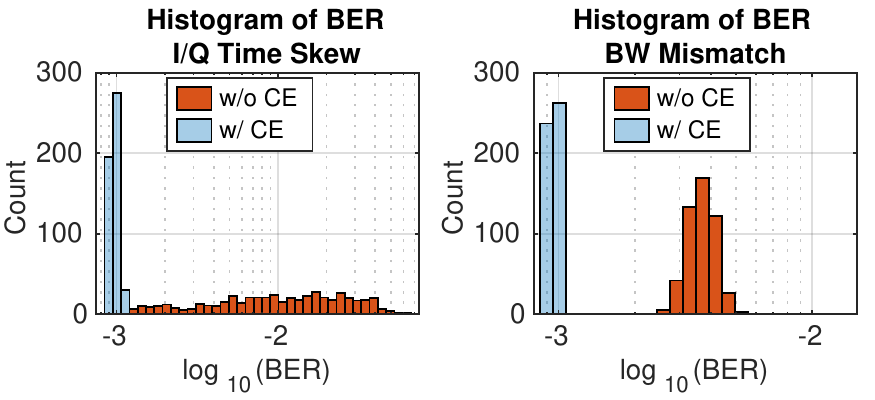}
\caption{\label{f:hist2}Histogram of the BER for 500 random cases with
  and without the CE for a reference BER of $\sim 1\times 10^{-3}$. Left:
  I/Q time skew (only). Right: BW mismatch (only).  See simulation
  parameters in Table~\ref{t:sim_parameters}.}
\end{figure}

Each Montecarlo test consists of 500 cases where the impairment
parameters are obtained from a UDRV random number
generator. Figs.~\ref{f:hist1} and~\ref{f:hist2} show the histograms
of the BER for the receiver with and without the CE in the presence of
sampling phase errors, gain errors, I/Q time skew, and BW mismatches.
Only one effect is exercised in each case.  Results for sampling phase
and gain errors uniformly distributed in the interval
$\delta_m^{(i)}\in [\pm 0.075]T$ and
$\Delta_{\gamma^{(i)}_m} \in [\pm 0.15]$ (see \eqref{eq:gain_error}),
respectively, are depicted in Fig. \ref{f:hist1}, whereas
Fig. \ref{f:hist2} shows results for random BW mismatches (see
\eqref{eq:B}) and I/Q time skews uniformly distributed in the interval
$\Delta_{B^{(i)}_m} \in [\pm 0.075]B_0$ and
$\tau_H,\tau_V\in [\pm 0.075]T$, respectively.  For all the evaluated
cases the proposed compensation technique is able to mitigate the
impact of all the impairments on the performance of the receiver when
they are exercised separately\footnote{DC offsets mismatch
  compensation has also been verified with similar performance
  improvement~\cite{solis_background_2020}.}.  Moreover, a BER
improvement of up to $10\times$ can be achieved with this proposal.
In particular, notice that the serious impact on the receiver
performance of the I/Q time skew values of
Table~\ref{t:sim_parameters} is practically eliminated by the proposed
CE with $L_g=7$ taps.

\begin{figure}[!t]
\centering
\includegraphics[width=.8\columnwidth]{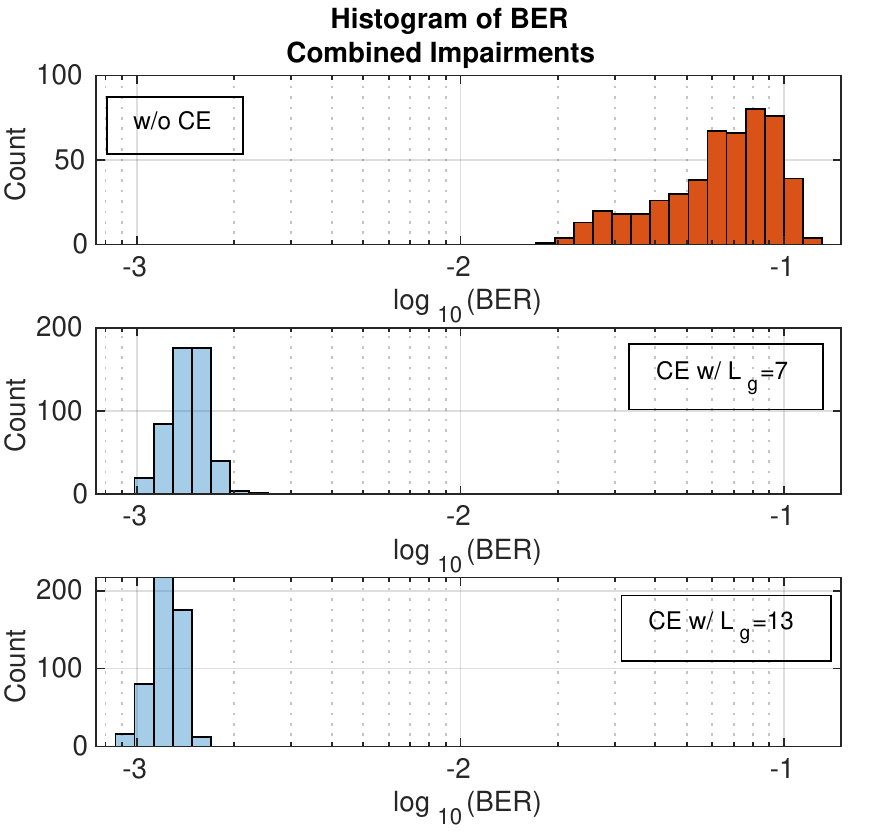}
\caption{\label{f:hist3}Histogram of the BER for 500 random cases with
  combined impairments as defined in Table~\ref{t:sim_parameters}.
Reference BER of $\sim 1\times 10^{-3}$. Top: without CE. Middle: CE w/$L_g=7$ taps.
Bottom: CE w/$L_g=13$ taps.}
\end{figure}

BER histograms for the receiver with and without the CE in the
presence of the combined effects are shown in Fig.~\ref{f:hist3}.
Results of 500 cases with random gain errors, sampling phase errors,
I/Q time skews, BW mismatches, and DC offsets as defined in
Table~\ref{t:sim_parameters}, are presented.  Fig.~\ref{f:hist3} also
depicts the performance of the CE with $L_g=13$ taps.  Without CE, a
severe degradation on the receiver performance as a consequence of the
combined effects of the TI-ADC mismatches is observed.  However, note
that the CE is able to compensate the impact of all combined
impairments improving the BER in some cases by almost $100$ times.
Moreover, note that a slight performance improvement can be achieved
increasing the number of taps $L_g$ from 7 to 13.

\begin{figure}[t]
\centering
\includegraphics[width=1.\columnwidth]{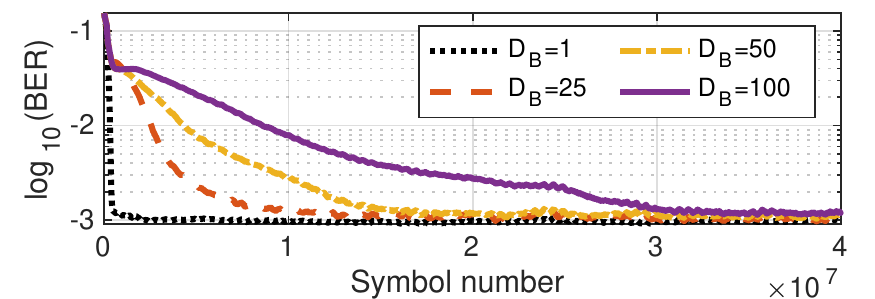}
\caption{\label{f:ber_conv}Convergence of the CE in the presence of
  combined impairments for different block decimation factors $D_B$
  with $N=8192$.}
\end{figure}

In multi-gigabit transceivers, the impairments of the AFE and TI-ADCs
change very slowly over time, as mentioned in
Section~\ref{s:bp_formulation}.  Hence, decimation can be applied
since the coefficient updates given by~\eqref{eq:lmsg2}
and~\eqref{eq:tildeo} do not need to be made at full rate.  In ultra
high-speed transceiver implementations (e.g., for optical coherent
communication), block processing and frequency domain equalization
based on the Fast Fourier Transform (FFT) are widely
used~\cite{morero_design_2016}.  Therefore, we propose to update the
CE performing block decimation over the error samples. The procedure
is detailed as follows.  Let $N$ be the block size in samples to be
used for implementing the EBP.  Define $D_B$ as the block decimation
factor.  In this way, the CE is updated using only one block of $N$
consecutive samples of the oversampled slicer error~\eqref{eq:oe}
every $D_B$ blocks, i.e.,
\begin{equation}
  e^{(i)}[k ND_B+n],\quad n=0,1,\cdots,N-1,\forall k\ ,
\end{equation}
with $k$ integer. By using this approach, Fig.~\ref{f:ber_conv} shows
an example of the temporal evolution of the BER in the presence of
combined impairments according to Table~\ref{t:sim_parameters} for
different values of $D_B$ with $N=8192$.
A moving average filter of length $10$ has been
used to process the instantaneous BER.  Gear shifting is used to
reduce the steady-state MSE and speedup the convergence of the
algorithm.  We highlight that the impact on the resulting BER is
negligible when block decimation is applied.  Therefore,
its adoption will drastically reduce the implementation complexity.

\subsection{Mixed-Signal Calibration of TI-ADC with Highly Interleaved
  Architectures}
\label{ss:high_interleaved_sim}
\begin{figure}
\centering
\includegraphics[width=1.\columnwidth]{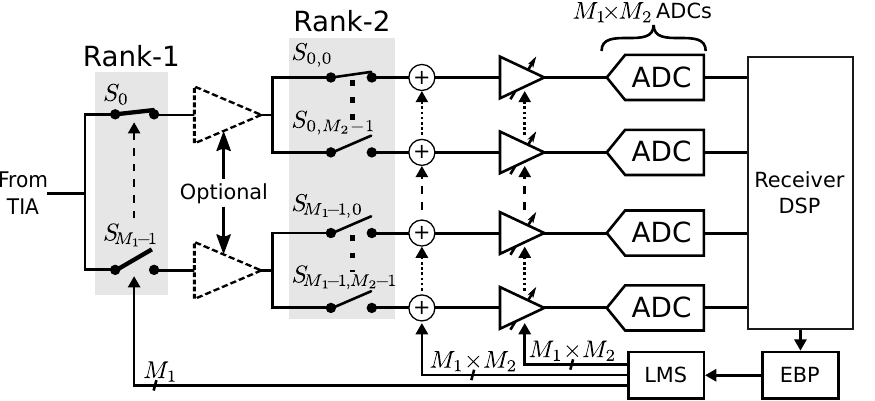}
\caption{\label{f:sch_hier_tiadc} Example of application of the
  proposed backpropagation based mixed-signal calibration in a typical
  two-rank hierarchical TI-ADC. For simplicity the details for only
  one lane are depicted.}
\end{figure}

\begin{figure}[t]
  \centering
  \includegraphics[width=.9\columnwidth]{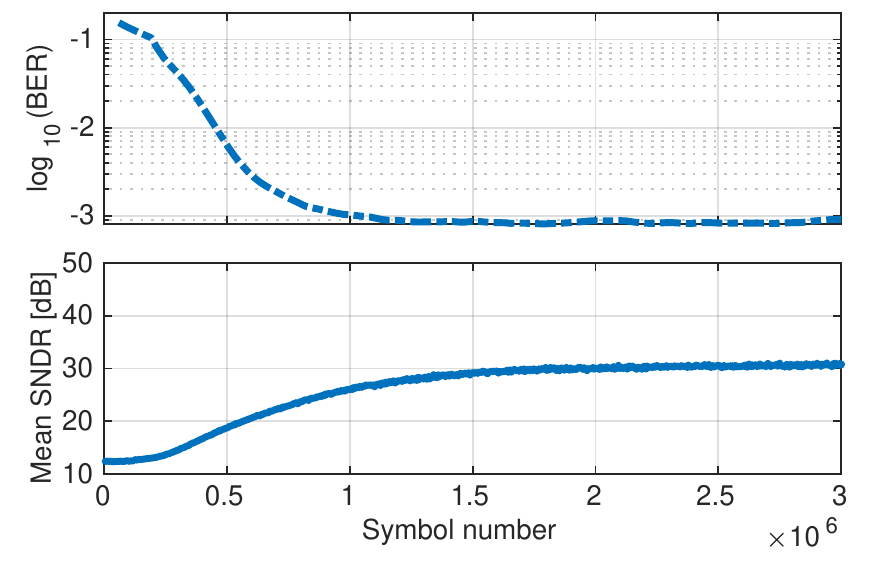}
  \caption{\label{f:ber_evo_jitter} BER and SNDR evolution in a
    hierarchical TI-ADC based DP optical coherent receiver with the
    backpropagation based mixed-signal calibration in the presence
    combined impairments. $M_1=16$ and $M_2=8$.}
\end{figure}

The mixed-signal approach described in Section~\ref{s:mixed} is
investigated considering a TI-ADC architecture typically used in high
speed receivers.  In such applications, a hierarchical TI-ADC achieves
the ultra high-speed and the best power
efficiency~\cite{reyes_energy-efficient_2019, kull_24-72-gs/s_2018,
  kim_161-mw_2020}. Successive Approximation Register (SAR) ADCs are commonly
used in this architecture due to their power efficiency at the
required sampling rate and resolution.
The hierarchical TI-ADC organizes the T\&H into
two or more ranks with a large number of sub-ADCs.  In this way the
requirements for the generation and synchronization of the sampling
clocks are relaxed.  Additionally, a massive number of sub ADCs with
low sampling rate and high power efficiency can be used.  Furthermore,
the impact on the input bandwidth is reduced in contrast to T\&H with
direct sampling~\cite{greshishchev_40_2010}.  An example with two
ranks, including the calibration elements required by our proposal in
this variant is depicted in Fig.~\ref{f:sch_hier_tiadc}.  Rank 1
comprises $M_1$ switches each of which feeds $M_2$ T\&H stages of rank
2.  Then, $M_1\times M_2$ sub ADCs are used to digitize the input
signal.    In order to evaluate the
performance of the mixed-signal calibration of Section~\ref{s:mixed},
we model a hierarchical TI-ADC with $M_1=16$ and $M_2=8$ (i.e.,
$M_1\times M_2=128$ sub ADCs).  Clock signals with \SI{100}{\fs} RMS
white-noise jitter are considered in this simulation. 
We emphasize that the sampling phases of the $M_1$ switches in the 
first rank, and the $M_1\times M_2$ gains and offsets of the 
sub-ADCs are adjusted with this approach.

The temporal evolution of the BER and the mean
Signal-to-Noise-and-Distortion-Ratio (SNDR) is shown in
Fig.~\ref{f:ber_evo_jitter}. A \SI{54}{\GHz} input tone is 
used in this measurement.
Since a larger number of converters is
used (i.e., 128 vs 16), a slightly slower convergence is observed with
respect to previous simulation.  Nonetheless, the impact of the
mismatches in a hierarchical TI-ADC performance is mitigated with the
proposed backpropagation based mixed-signal calibration.  In
particular, note that the SNDR can be improved from \SI{\sim12}{\dB}
to \SI{\sim30}{\dB} by using the proposed background technique.

{
Figure~\ref{f:sim_fft_comparison} shows a comparison of the FFTs 
pre and post calibration with a \SI{54}{\GHz} sinusoidal input. 
Spectrum is generated from $2^{13}$ samples from the ADC of
polarization $H$, component $I$. Measured amplitudes 
are normalized to Full-Scale (FS)~\cite{kester_data_2005}.
Without calibration the mismatches introduce a large number of 
spurs with significant amplitude all across the spectrum.
After the calibration, the spurs are greatly reduced.
Hence, the SNDR is improved from \SI{12.3}{\dB FS} to 
\SI{30.6}{\dB FS}. The Spurious-Free Dynamic Range (SFDR) is also boosted
from \SI{23.6}{\dB FS} to \SI{51.2}{\dB FS}.

\begin{figure}
    \centering
    \includegraphics[width=0.9\columnwidth]{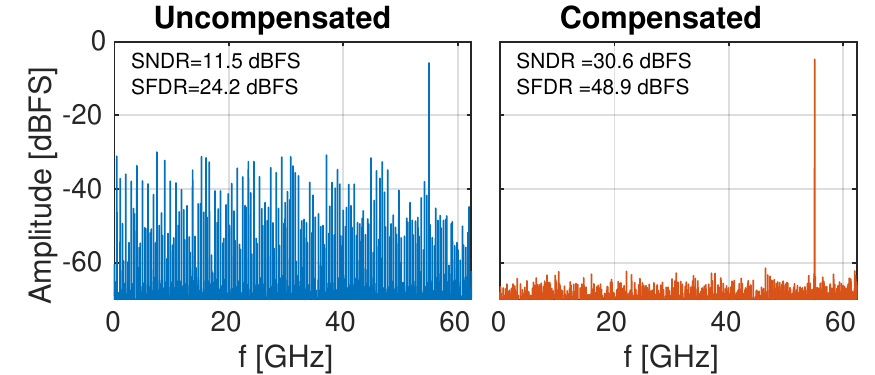}
    \caption{Spectrum comparison for a sinusoidal input at \SI{55}{\GHz}
    after and before applying the calibration of 
    Fig.~\ref{f:ber_evo_jitter}. $2^{13}$ samples are used to
    generate the plots.}
    \label{f:sim_fft_comparison}
\end{figure}

The SNDR and the SFDR with and without the
calibration are shown in Fig.~\ref{f:sim_sndr_sfdr}. Prior to calibrate, the
SNDR and SFDR are below \SI{20}{\dB FS} and \SI{31}{\dB FS}, respectively.
With the calibration enabled, the SNDR remains above \SI{40}{\dB FS}
until \SI{10}{\GHz}. At higher frequencies the SNDR is limited by the jitter asymptote. The SFDR is above \SI{50}{\dB FS} for the 
evaluated frequencies. Thus, a significant improvement in the ADC is
obtained with the proposed technique.

\begin{figure}[t]
  \centering
  \includegraphics[width=.9\columnwidth]{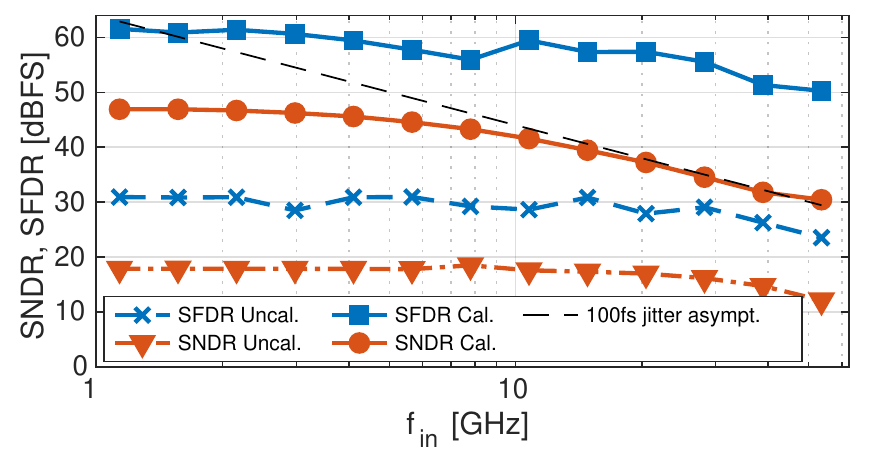}
\caption{\label{f:sim_sndr_sfdr}SNDR and SFDR versus input frequency
  with and without calibration.}
\end{figure}

A comparison with other calibration techniques is summarized
in Table~\ref{t:comparison}. We emphasize that our proposal operates in
background, does not require reference channels, a particular type of input
signal, or a modified sampling sequence. Furthermore, the calibration can
be applied in either analog or digital domain. In addition, the calibration
engine is not limited to estimate and correct only sampling time errors, 
but also gain, DC offset, bandwidth, and the I/Q time skew present in
coherent communication systems. Moreover, the technique is able to adjust
the errors simultaneously (i.e., a calibration sequence is
not needed).

\begin{table}[t]
\centering
\caption{Summary and comparison with other state-of-the-art techniques.}
\label{t:comparison}
  \begin{tabular}{l|c|c|c|c|c}
    \hline
    {\textbf{Features}}      & 
    \begin{tabular}{@{}c@{}}\textbf{2020} \\ \cite{ali_12-b_2020}\end{tabular} &
    \begin{tabular}{@{}c@{}}\textbf{2020} \\ \cite{guo_16-gss_2020} \end{tabular} &
    \begin{tabular}{@{}c@{}}\textbf{2017} \\ \cite{le_duc_fully_2017}\end{tabular} &
    \begin{tabular}{@{}c@{}}\textbf{2019} \\ \cite{salib_high-precision_2019}\end{tabular} & 
    \begin{tabular}{@{}c@{}}\textbf{This} \\ \textbf{work}\end{tabular}
    \\ \hline 
    Background             & Yes & Yes & Yes  & Yes & \textbf{Yes} \\ 
    Blind                  & Yes & Yes & Yes  & Yes & \textbf{Yes} \\ 
    Reference channels     & No & Yes & No & No & \textbf{No}    \\ 
    Dither injection       & Yes & No & No & No & \textbf{No}\\ 
    Regular samp. seq.     & Yes & Yes & Yes & No & \textbf{Yes}\\ 
    Number of interleaves  &  8  & 7/8 & 4 & 8 & \textbf{4 to 16} \\ 
    Cal. scheme              & Dig  & Dig & Dig & Dig & \textbf{A/Dig} \\ 
    Simultaneous conv.     & Yes & No & No & N/A & \textbf{Yes}     \\ \hline
  \end{tabular}
\end{table}
}
\section{Experimental Results}
\label{s:experimental}

We demonstrate the benefits of our proposal using a digital
communication platform especially designed to evaluate TI-ADC mismatch
calibration techniques.  The platform allows the capabilities of our
proposal to be evaluated in communication links using several
different modulation schemes.  Some key differences with more
traditional TI-ADC evaluation platforms described in the literature
are the following:
\begin{enumerate}
\item Test signals are not limited to sinusoids, but they also include
  realistic communications signals.
\item \label{it:2}Characterization of the TI-ADC output is not limited
  to spectral estimation, but it additionally incorporates a complete
  communications receiver DSP.
\item The samples generated by the TI-ADC are not decimated as is
  usually done in experiments described in the literature.  This is
  important in our experiments, since the receiver DSP mentioned
  in~\ref{it:2}) requires contiguous samples and it cannot operate
  properly with decimated samples.
\end{enumerate}

The test chip has programmable delay cells that allow the mixed-signal
sampling phase calibration variant of our proposal to be tested too.
We highlight that the following results can be ported to the
high-speed optical coherent transceiver application scenario exercised
in Section~\ref{s:sim_res} since both the receiver and EBP blocks are
the same as those used in previous simulations.

\subsection{High-Speed Time-Interleaved SAR ADC}
\label{ss:tchip3}
\begin{figure}[t]
\centering
\includegraphics[width=.95\columnwidth]{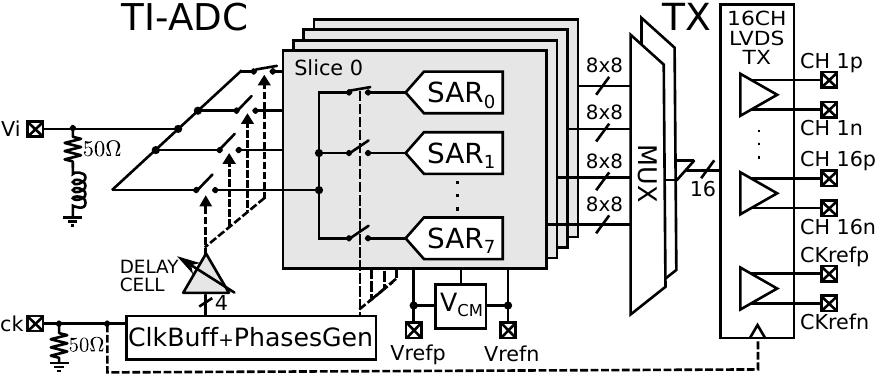}
\caption{\label{f:tchip3_top}Top architecture block diagram of the
  TI-ADC test chip. Analog input and clock signals are fully
  differential.}
\end{figure}

\begin{figure}[t]
\centering
\includegraphics[width=0.4\columnwidth]{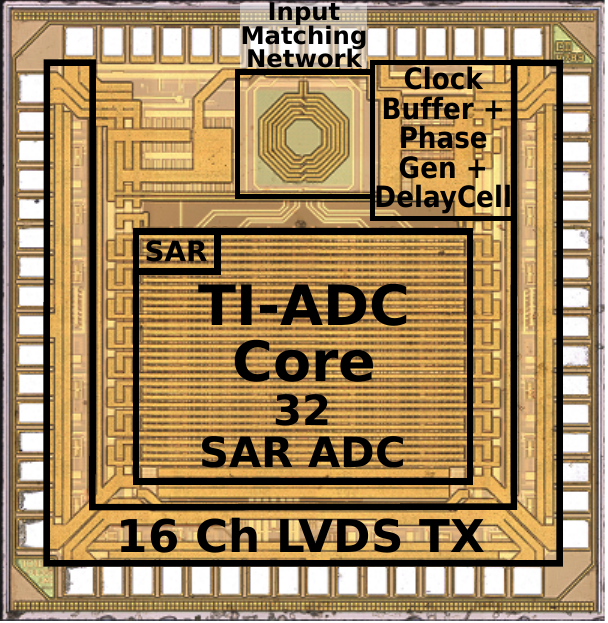}
\caption{\label{f:tchip3_photo}Photograph of the TI-ADC test chip used
  in the receiver. The resolution of the converter is \SI{8}{b}, the
  nominal sampling rate is \SI{4}{\giga\sample\per\s}, and it is
  fabricated in a \SI{130}{\nm} CMOS process.}
\end{figure}

A block diagram of the \SI{4}{\giga\sample\per\s}, \SI{8}{b} TI-ADC
architecture~\cite{solis_4gss_2021} is shown in
Fig.~\ref{f:tchip3_top}.  A photograph of the test chip, which is
fabricated in a~\SI{130}{\nm} CMOS process from Global Foundries, is
shown in Fig.~\ref{f:tchip3_photo}.  The TI-ADC core area is
\SI{1}{\mm\squared} whereas the total area of the chip is
\SI{2}{\mm}$\times$\SI{2}{\mm}.  The design is composed of an input
matching network, a hierarchical, non-buffered T\&H, a TI-ADC core, a
high-speed Low-Voltage Differential Signaling (LVDS) interface, and a
clock sampling phase generator.  The input matching network includes
\SI{50}{\ohm} resistors and an \SI{8}{\nano\henry} inductor to enhance
the tracking bandwidth.  The T\&H spans two sampling hierarchies.  The
first sampling hierarchy has $M_1=4$ switches whereas the second has
$M_2=8$, which are included in the sub-ADCs. {Since this T\&H
  architecture avoids the use of sampling buffers, the noise sources
  in the signal path are minimized, and power consumption is
  reduced~\cite{reyes_energy-efficient_2019}.  As mentioned in
  Section~\ref{ss:high_interleaved_sim}, in a hierarchical T\&H
  architecture the sampling time errors mainly depend on the clock
  signals of the first hierarchy.  Then, in order to adjust the
  sampling phases of the first hierarchy, capacitor-based programmable
  delay cells are included.}  The maximum calibration range of the
delay cells is approximately \SI{\pm50}{\ps}, with a minimum
calibration step of \SI{260}{\fs}.  The core of the architecture is
comprised of an array of 32 power efficient asynchronous SAR ADCs
operating at \SI{125}{\msps} with \SI{8}{b} resolution.  A strongARM
comparator with on-chip DC offset calibration is used to quantize the
samples.  The common-mode voltage, $V_{CM}$ is generated on-chip as
the mean of the external reference voltages, $V_{refp}$ and
$V_{refn}$. The 256 digital outputs of the ADC core are
time-multiplexed and sent off-chip using a 16-channel LVDS driver.
The data is transferred without any decimation at \SI{32}{\giga
  b\per\s}.  An additional LVDS channel is used to transmit a clock
reference in order to achieve synchronization with the LVDS receiver.
To configure and control the different blocks in this architecture,
254 configuration registers are used.

The prototype chip achieves a peak ENOB of \SI{7.09}{\bit}
(\SI{5.47}{\bit} at Nyquist), \SI{1.3}{\GHz} bandwidth, and
\SI{93}{\mW} power consumption from a \SI{1.2}{\V} power supply. The
SAR ADC of each interleave achieves a Figure of Merit (FOM) of
\SI{123}{\femto\J\per conv-step}. The efficient interleaved
architecture allows the TI-ADC to achieve a peak FOM of
\SI{171}{\femto\J\per conv-step} (\SI{526}{\femto\J\per conv-step} at
Nyquist). The achieved efficiency and sampling rate are comparable to
designs where much more advanced CMOS process nodes are used.  For a
detailed description and additional measurements of the prototype
chip, please see~\cite{solis_4gss_2021}.

\subsection{Reconfigurable Experimental Platform}
\label{ss:setup}

\begin{figure}[t]
\centering
\includegraphics[width=0.95\columnwidth]{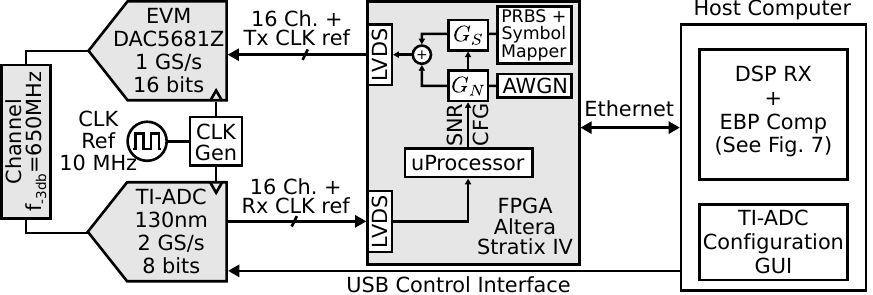}
\caption{\label{f:exp_setup}Experimental setup used to test the
  proposal.  A prototype TI-ADC with 8b and \SI{2}{\giga\sample\per\s}
  acquires the symbols synthesized by the DAC. To emulate the
  operation of the optical coherent system, several sets of samples
  are collected. Each set corresponds to a signal component and is
  obtained after setting the platform with different PRBS lengths and
  delay cell values.}
\end{figure}

\begin{figure}[t]
\centering
\includegraphics[width=.9\columnwidth]{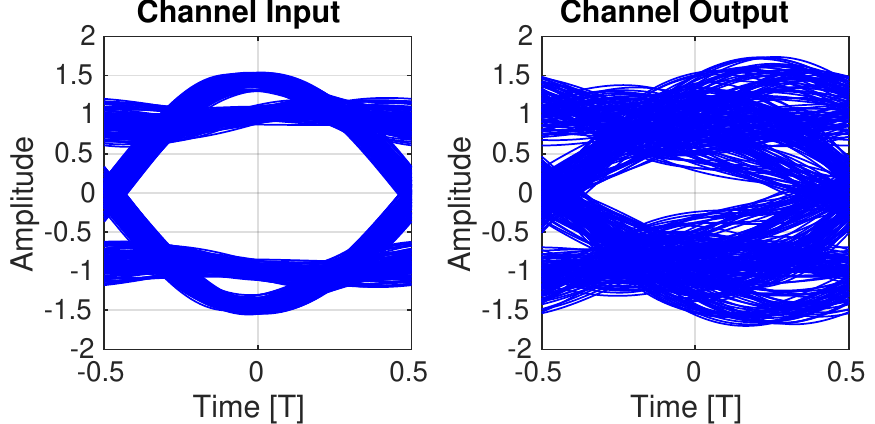}
\caption{\label{f:eyes}Measured eye diagrams at the input and output
  of the channel.}
\end{figure}

{
A block diagram of the experimental setup is shown in
Fig.~\ref{f:exp_setup}.  A high-performance Field-Programmable Gate
Array (FPGA)~\cite{altera_stratix_2014} is used to generate the
symbols to be transmitted.  The FPGA is also in charge of collecting
the samples from the ADC and sending them to the receiver DSP, which
is implemented on a host computer. The receiver architecture,
including the proposed compensation technique, has been introduced in
Fig.~\ref{f:f6} and simulated in Section~\ref{s:sim_res}.
Multiple Pseudo-Random Binary Sequences (PRBSs) with configurable
length and seed are generated in the FPGA.
The amplitude of the symbols and the Additive White Gaussian
Noise (AWGN) can be set through the coefficients $G_S$ 
and $G_N$, respectively. Then, we are able to evaluate different
Signal-To-Noise Ratio (SNR) scenarios.
The symbol with added noise is sent to a commercial,
16-bit Digital-to-Analog Converter (DAC)
board~\cite{texas_instruments_dac568181z82zevm_2008} using an LVDS
interface.  The DAC synthesizes the samples at
$1/T=\SI{1}{\giga\sample\per\s}$.  This sampling rate is adopted due
to limitations on the FPGA and DAC clocks.  The communication channel
is modeled as a low-pass filter with a \SI{-3}{\dB} cut-off frequency
of \SI{650}{\MHz}~\cite{mini-circuits_coaxial_2008}. Figure
\ref{f:eyes} shows the measured eye diagrams at the input and output
of the channel with Binary Phase Shift Keying (BPSK)
modulation. Notice that significant ISI is added by the channel.
Although not explicitly shown, the impact of the ISI is even more
significant for the higher order modulations used in the experiments,
such as 8-PAM/64-QAM and 16-PAM/256-QAM. This ISI is an
important part of the experiment since it enables the verification of
the backpropagation technique, as discussed later in this section.  On
the receiver side, the signal is acquired by the TI-ADC described in
Section~\ref{ss:tchip3}, operating at a sampling rate of
\SI{2}{\giga\sample\per\s} (i.e., an oversampling ratio of $T/T_s=2$
is used in the DSP blocks).  The clocks for both DAC and ADC are
generated from a single \SI{10}{\MHz} clock reference.  Therefore, the
frequency error due to the part per million tolerance of the
oscillators in the receiver and transmitter clocks is avoided.
}

{
As mentioned before, the acquired samples are sent to the FPGA by an
LVDS interface without applying any decimation.  
A set of $2^{18}$ consecutive samples (i.e., $2^{17}$ symbols) is
sent to the host computer in each
iteration.  Since the available experimental setup has one TI-ADC, a
suitable signal for the coherent receiver has to be assembled by
combining four independent measurements. This is done by collecting
one set of samples for each signal component. For a particular 
component of the complex signal, the platform is configured with
a unique PRBS length.  Furthermore, the configuration of the
delay cells is also changed for each component.
The post-processing routine on the computer incorporates a low
complexity CE with 4 sets of 15 independent coefficients for each
component, as described in Section~\ref{ss:dig_comp}, and the coherent
receiver used in simulations of Section~\ref{s:sim_res}. 
}

For the mixed-signal variant the CE is disabled, and the calibration
is performed adjusting the on-chip programmable delay cells according
to the computation performed by the EBP block, as described in
Section~\ref{s:mixed}.  The delay cells are updated after processing a
complete set of samples, prior to performing a new capture of samples
from the platform.

\subsection{Measurements}
\label{ss:measurement}

\begin{figure}[t]
\centering
\includegraphics[width=.9\columnwidth]{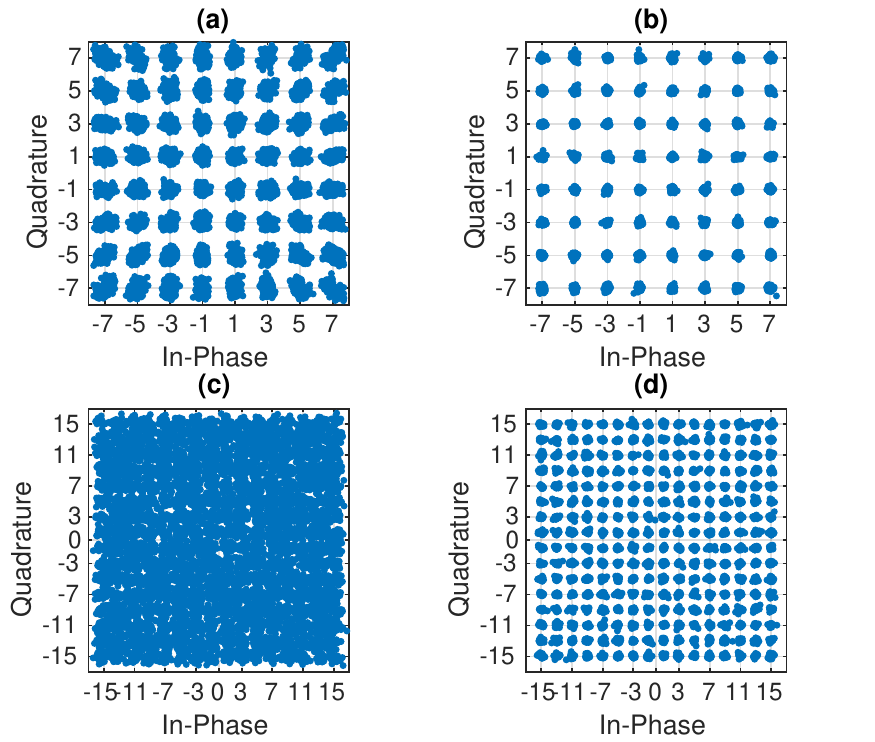}
\caption{\label{f:qam_const}
Received (a, b) 64-QAM, and (c, d) 256-QAM constellation diagrams in the presence of TI-ADC mismatches where the proposed compensation technique is (left) disabled and (right) enabled.
A noiseless channel is set to evaluate the impact of the TI-ADC mismatches on the receiver performance.}
\end{figure}

\begin{figure}[t]
\centering
\includegraphics{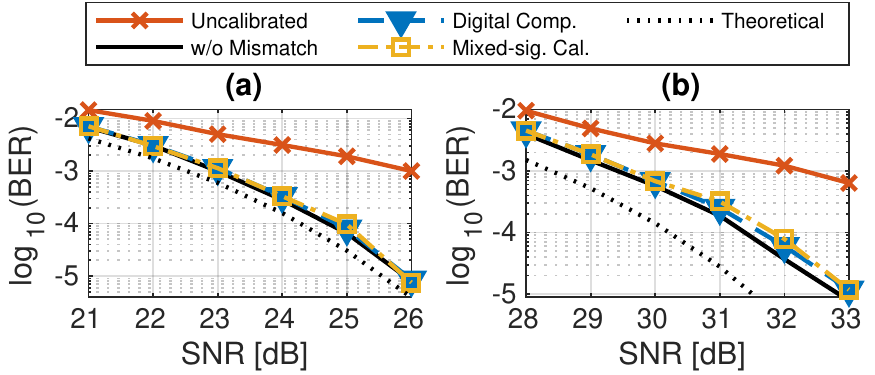}
\caption{\label{f:exp_ber_snr_qam}
Measured BER comparison for (a) 64-QAM, and (b) 256-QAM with and without the proposed calibration.
A mismatch of \SI{\pm4}{\percent}$T$ for (a), and \SI{\pm1}{\percent}$T$ for (b) is set using the delay cells.}
\end{figure}

In this section we present measurement results using the test chip and
the communication platform already described.  Since the test chip has
flexible timing calibration capabilities, we focus on the calibration
of sampling phase errors.  In addition, gain errors are also
compensated to demonstrate how this technique is able to
simultaneously compensate different types of mismatches.  Although
the DC offset mismatch causes severe degradation on the performance of
both TI-ADC and receiver, we do not consider such calibration here
since it is already calibrated on-chip~\cite{solis_4gss_2021}.

First, we illustrate the effectiveness of our proposal considering
only the impairments of the TI-ADC for a receiver using 64-QAM
and 256-QAM schemes.  Toward this end, a noiseless channel has been
set up.  The resulting constellation diagrams are shown in
Fig.~\ref{f:qam_const} for the aforementioned modulations.  In the
absence of compensation, the mismatch among the interleaves is large
enough to enlarge the constellation points considerably.  For a
256-QAM (see Fig.~\ref{f:qam_const}(b)) the degradation is such that
the symbols in the received constellation are not distinguishable.
With the proposed technique, a great improvement is observed for 
the modulations tested.

The comparison of the BER curves for the receiver with and without the
proposed technique is shown in Fig.~\ref{f:exp_ber_snr_qam}.  Both all
digital and mixed-signal variants are evaluated using 64-QAM
and 256-QAM schemes.  The performance of the receiver is severely
affected when the TI-ADC mismatch is not mitigated.  A sampling phase
error of \SI{4}{\percent} has been set for
Fig.~\ref{f:exp_ber_snr_qam}(a),
whereas \SI{1}{\percent} is set for
Fig.~\ref{f:exp_ber_snr_qam}(b).  Setting a larger sampling phase
error for QAM-256 would incur in issues related to the convergence of
the receiver.  A considerably high SNR penalty of
\SI{3}{\dB} is measured for a 64-QAM modulation at a BER of
\num{1e-3}.  A similar penalty can be observed for a 256-QAM scheme,
although the mismatch in this case is much smaller than in the
previous case.  After enabling the proposed technique, the performance
of the receiver is restored to almost replicate the case without
mismatch with both implementation variants.  This result indicates
that our proposal is able to nearly eliminate the receiver penalty
introduced by the mismatches of the TI-ADC.

\begin{figure}[t]
\centering
\includegraphics[width=.8\columnwidth]{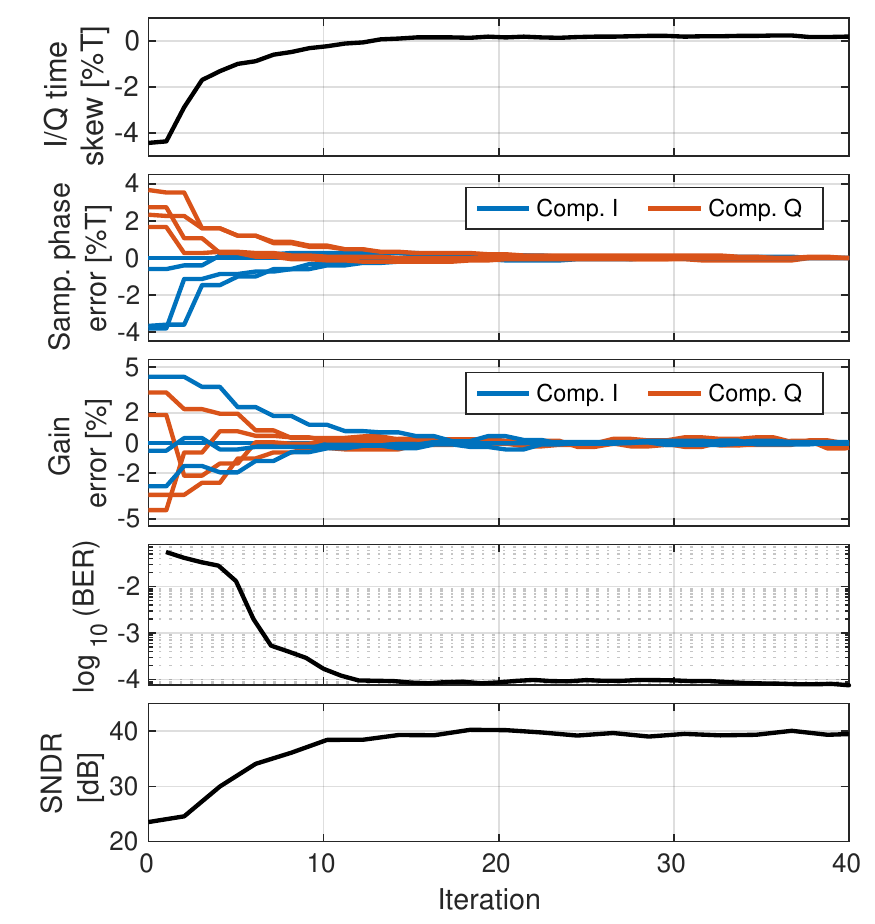}
\caption{\label{f:exp_taps_convergence} {Example of calibration
    convergence of I/Q channels of horizontal polarization in a 64-QAM
    scheme with an SNR of \SI{\sim25}{\dB}. The subfigures from top to
    bottom are: I/Q time skew, sampling phase error, gain error, BER,
    and SNDR.}}
\end{figure}

{
An example of the convergence for the mixed-signal calibration using a
64-QAM scheme is shown in Fig.~\ref{f:exp_taps_convergence}.  Reported
measurements are taken from polarization $H$.  I/Q time skew has been
introduced in each component by initially shifting all the delay cells
by the same amount.  An initial I/Q time skew of \SI{\sim4}{\ps} has
been set in this test.  
Gain mismatch is added to the samples in the post processing to
exacerbate its effect on the receiver performance.
The gain error is also adjusted in the post processing routine using
the EBP block.  The delay cells
and post processed gain mismatch are updated according to the
estimation obtained in each iteration of the test.  
We emphasize that the proposal is able
to mitigate the I/Q time skew by tuning all the delay cells present
in the test chip. In this
measurement the target BER is \num{1e-4}.  The instantaneous BER is
computed after performing a calibration step, and then processed by a
moving average filter of size 6. The target BER is reached after
12 iterations, which implies the processing of
\num{1.5e6} received symbols or \SI{1.5}{\ms} in a
\SI{1}{\giga\baud} link.  Considering a high-speed optical coherent
application, such as the \SI{96}{\giga\baud} link exercised in
Section~\ref{s:sim_res}, the convergence would be achieved after
\SI{16}{\micro\s}.
The SNDR is measured applying a \SI{\sim500}{\MHz}
sinusoidal input and setting the delay cells and post processed
gain mismatch according to their evolution in the experiment just
described.  As a result, the SNDR is improved from \SI{24}{\dB} to
\SI{40}{\dB}. We highlight that in Fig.~\ref{f:exp_taps_convergence}
the estimators of \textit{all} the errors for \textit{all} the
interleaves are adjusted \textit{simultaneously}.  Hence, a
sequence of calibration steps (e.g., calibrate the sampling phase
first, then the gain) is not needed, thus improving the speed of
convergence.  Finally, since the proposed technique runs in
background, the mismatches are reduced concurrently with the
convergence of the receiver.
}

\begin{figure}[t]
\centering
\includegraphics[width=.9\columnwidth]{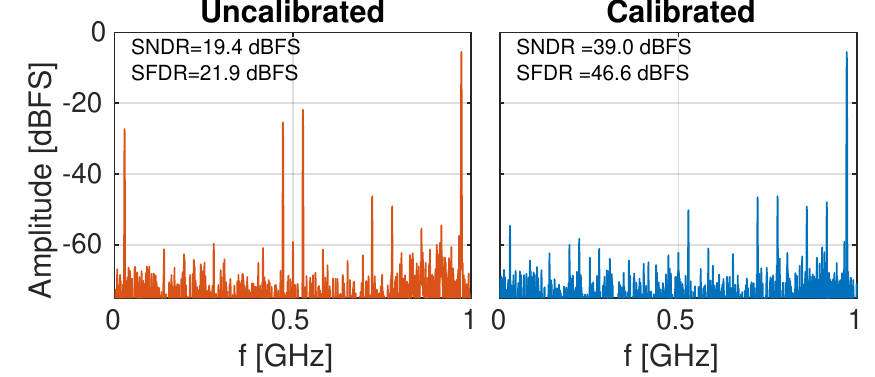}
\caption{\label{f:exp_fft_prepost}Spectrum comparison for a sinusoidal
  input signal at \SI{972}{\MHz} with \SI{\pm4}{\percent}$T$ of
  mismatch in the sampling phase and \SI{\pm5}{\percent} of gain
  mismatch with respect to the unity gain. 
  The spectra are generated from $2^{13}$ samples.}
\end{figure}

The spectrum comparison for a sinusoidal input at \SI{972}{\MHz} pre
and post calibration is shown in Fig.~\ref{f:exp_fft_prepost}.
Samples from a single channel, (i.e., from polarization $H$, component
$I$) have been used to generate the spectra.  \SI{\pm4}{\percent}$T$
of mismatch in the sampling phase and \SI{\pm5}{\percent} of gain
mismatch with respect to the unity gain have been included. 
Since the mismatches in the first sampling hierarchy are predominant,
we observe $M_1-1=3$ spurs with high amplitude in the spectrum.
Notice that the spurs caused by the mismatches among the interleaves
seriously degrade both the SNDR and SFDR to \SI{19.4}{\dB FS} and \SI{21.9}{\dB FS}, respectively.
After applying the proposed technique, the performance of the TI-ADC
is boosted to \SI{39}{\dB FS} and \SI{46.6}{\dB FS}, for SNDR and
SFDR, respectively.


\begin{figure}[t]
  \centering
  \includegraphics[width=.9\columnwidth]{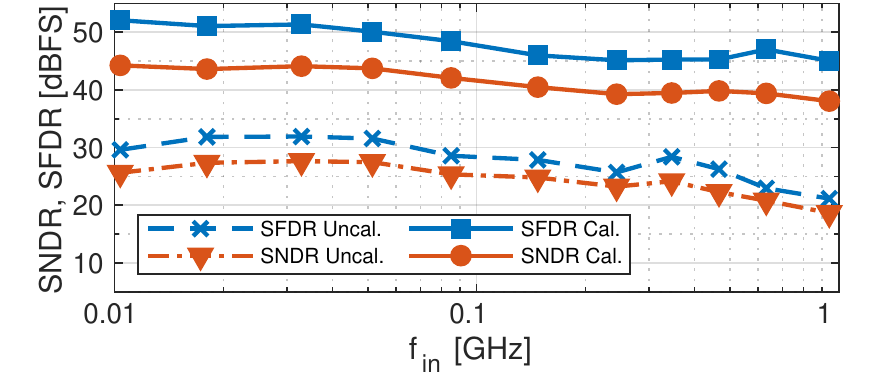}
\caption{\label{f:exp_sndr_sfdr}SNDR and SFDR versus input frequency
  with and without compensation.  A mismatch of \SI{\pm4}{\percent}$T$
  in the sampling phase and \SI{\pm5}{\percent} of the unity gain are
  set in this measurement.}
\end{figure}

The measurement of the SNDR and SFDR as a function of the input
frequency with and without the proposed technique is shown in
Fig.\ref{f:exp_sndr_sfdr}. Mismatches of sampling phase and gain are
distributed as in the previous measurement. Without any calibration,
the SNDR and SFDR are below \SI{28}{\dB FS} and \SI{32}{\dB FS},
respectively, for all the frequency range.  After applying the
backpropagation-based calibration the performance of the converter is
significantly improved.  For all the Nyquist range the SNDR is higher
than \SI{39}{\dB FS} while the SFDR remains above \SI{46}{\dB FS} in
the same frequency range.  At least \SI{15}{\dB} of improvement in
both measurements is achieved with this proposal.

\section{Conclusions}
\label{s:conclusion}
A novel background calibration technique for TI-ADC mismatches based
on the backpropagation algorithm has been introduced in this paper.
The characteristics of the backpropagation algorithm are exploited in
a digital communication receiver application, where the algorithm is
used to generate a suitable error signal, which is processed to
effectively mitigate the mismatches of a high-speed TI-ADC. The
technique can be extended to compensate impairments of the entire AFE
(e.g., I/Q time skew).  This proposal can be implemented either with a
fully digital or a mixed-signal approach.  Simulations performed in an
application example with a DSP-based, DP optical coherent receiver
have shown a fast, robust and almost ideal compensation/calibration of
different TI-ADC mismatches.  Sampling time, gain, offset, and
bandwidth mismatches as well as I/Q time skew errors have been
exercised both individually and combined.  Measurements have been
performed using an emulation platform based on an 8 bit, up to
\SI{4}{\gsps} TI-ADC test chip.  We have shown that the degradation in
the receiver performance is highly mitigated with this proposal for
64-QAM and 256-QAM schemes.  Moreover, an SNDR improvement of
at least \SI{\sim15}{\dB} is measured for all the Nyquist range. We
highlight that this proposal is able to compensate mismatches of
several types simultaneously (i.e., a sequence of calibration is not
needed).  Hardware complexity is minimized using decimation and serial
processing in the backpropagation blocks.  As the technique runs in
background, the proposed technique is able to track parameter
variations caused by temperature, voltage, aging, etc., without
operational interruptions.
\appendix
\section*{Appendix A: TI-ADC Model}
\label{app:tiadc_model}
\begin{figure}
  \centering
  \includegraphics[width=1.\columnwidth]{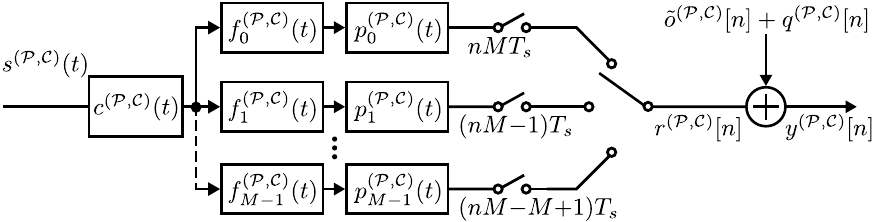}
  \caption{\label{f:fig2} Modified model of the analog front-end and
    TI-ADC for polarization $\mathcal P \in\{H,V\}$ and component
    $\mathcal C \in\{I,Q\}$ in a DP coherent optical receiver.}
\end{figure}

Next we review the model of the TI-ADC with impairments used in this
paper (see Fig. \ref{f:fig1}). The effects of the sampling time errors
$\delta_m^{({\mathcal P},{\mathcal C})}$ and gain errors
$\gamma_m^{({\mathcal P},{\mathcal C})}$ can be modeled by analog
interpolation filters with impulse responses
$p_m^{({\mathcal P},{\mathcal C})}(t)$ followed by ideal
sampling~\cite{luna_compensation_2006,agazzi_90_2008}, as depicted in
Fig.~\ref{f:fig2}.  The digitized high-frequency samples can be
written as
\begin{equation}
\label{eq:eq1}
y^{({\mathcal P},{\mathcal C})}[n]=r^{({\mathcal P},{\mathcal C})}[n]+{\tilde o}^{({\mathcal P},{\mathcal C})}[n]+q^{({\mathcal P},{\mathcal C})}[n],
\end{equation}
where $r^{({\mathcal P},{\mathcal C})}[n]$ is the signal component
provided by the $M$-channel TI-ADC, and
$q^{({\mathcal P},{\mathcal C})}[n]$ is the quantization noise.

\begin{figure}
  \centering
  \includegraphics[width=.9\columnwidth]{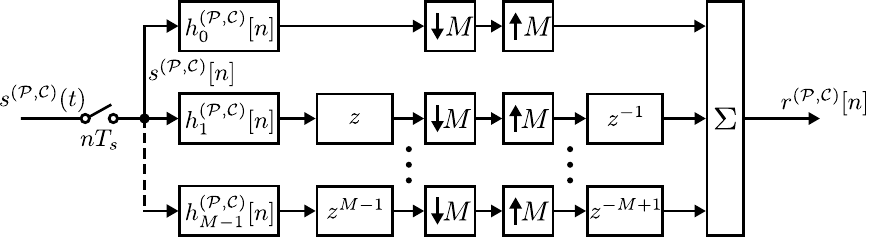}
  \caption{\label{f:fig3} Equivalent discrete-time model of the analog
    front-end and TI-ADC system with impairments for the signal
    component given by \eqref{eq:eq4} (i.e., without DC offsets and
    quantization noise) for polarization $\mathcal P \in\{H,V\}$ and
    component $\mathcal C \in\{I,Q\}$.}
\end{figure}

The total impulse response of the $m$-th interleaved channel is
defined as
\begin{equation}
\label{eq:eq2}
h_m^{({\mathcal P},{\mathcal C})}(t)=c^{({\mathcal P},{\mathcal C})}(t)\otimes f_m^{({\mathcal P},{\mathcal C})}(t) \otimes p_m^{({\mathcal P},{\mathcal C})}(t),
\end{equation}
where $m=0,\cdots, M-1$ and $\otimes$ is the convolution operator.
Let $H_m^{({\mathcal P},{\mathcal C})}(j\omega)$ and
$S^{({\mathcal P},{\mathcal C})}(j\omega)$ be the Fourier Transforms
(FTs) of $h_m^{({\mathcal P},{\mathcal C})}(t)$ and
$s^{({\mathcal P},{\mathcal C})}(t)$, respectively.  The spectral
shaping commonly used in digital communication systems results in
$|S^{({\mathcal P},{\mathcal C})}(j\omega)|\approx 0$ for
$|\omega|\ge \pi/T_s$.  Then, the analog filtering of
Fig.~\ref{f:fig2} can be replaced (assuming
$|H_m^{({\mathcal P},{\mathcal C})}(j\omega)|\approx 0$ for
$|\omega|\ge \pi/T_s$) by a real discrete-time model, as depicted in
Fig. \ref{f:fig3}, resulting
\begin{equation}
\label{eq:eq3}
h^{({\mathcal P},{\mathcal C})}_m[n]=T_sh_m^{({\mathcal P},{\mathcal C})}(nT_s),\quad m=0,\cdots,M-1.
\end{equation}
Therefore, it can be shown that the digitized high-frequency signal
can be expressed as:
\begin{equation}
\label{eq:eq4}
r^{({\mathcal P},{\mathcal C})}[n]=\sum_{l} {\tilde h}^{({\mathcal P},{\mathcal C})}_n[l] s^{({\mathcal P},{\mathcal C})}[n-l],
\end{equation}
where
$s^{({\mathcal P},{\mathcal C})}[n]=s^{({\mathcal P},{\mathcal
    C})}(nT_s)$ and ${\tilde h}^{({\mathcal P},{\mathcal C})}_n[l]$ is
the impulse response of a time-varying filter, which is an
$M$-periodic sequence such
${\tilde h}^{({\mathcal P},{\mathcal C})}_n[l]={\tilde h}^{({\mathcal
    P},{\mathcal C})}_{n+M}[l]$, and defined by
\begin{equation}
  \label{eq:eq5}
  {\tilde h}^{({\mathcal P},{\mathcal C})}_n[l]={h}^{({\mathcal P},{\mathcal C})}_n[l], \quad n=0,\cdots,M-1,
  \forall l,
\end{equation}
with ${h}^{({\mathcal P},{\mathcal C})}_n[l]$ given by
\eqref{eq:eq3}\footnote{See \cite{saleem_adaptive_2010} and references
  therein for more details about this formulation.}.  We highlight
that the impact of both the AFE impairments and the $M$-channel TI-ADC
mismatches are included in \eqref{eq:eq4}.  Finally, the digitized
high-frequency sequence is obtained by replacing \eqref{eq:eq4} in
\eqref{eq:eq1},
\begin{align}
  \nonumber
  y^{({\mathcal P},{\mathcal C})}[n]=&\sum_{l} {\tilde h}^{({\mathcal P},{\mathcal C})}_n[l] s^{({\mathcal P},{\mathcal C})}[n-l]+{\tilde o}^{({\mathcal P},{\mathcal C})}[n]+\\
  \label{eq:eq1bb}
              &q^{({\mathcal P},{\mathcal C})}[n].
\end{align}

\section*{Appendix B: DSP MIMO Backpropagation Details}
\label{app:backprop}
In this Appendix we derive an expression for the stochastic gradient
of the squared error defined by~\eqref{eq:grad}.  The total squared
error~\eqref{eq:eT} is
\begin{equation}
  \label{eq:eT2}
  {\mathcal E}_k=\sum_{j=1}^4 \left|e_k^{(j)}\right|^2=\sum_{j=1}^4
  \left(u_k^{(j)}- \hat a_k^{(j)}\right)^2,
\end{equation}
where $u_k^{(j)}$ is given by~\eqref{eq:u1}.  We define the
\emph{average} squared error as
\begin{equation}
  \label{eq:mse}
  \overline {\mathcal E}_N=\frac{1}{2N+1}\sum_{k=-N}^{N}\sum_{j=1}^4
  \left(u_k^{(j)}-\hat a_k^{(j)}\right)^2.
\end{equation}

The derivative of $\overline {\mathcal E}_N$ with respect to
$g^{(i_0)}_{m_0}[l_0]$ can be defined as
\begin{equation}
  \label{eq:dEdg}
  \frac{\partial { \overline {\mathcal E}_N}}{\partial g^{(i_0)}_{m_0}[l_0]}=\frac{2}{2N+1}\sum_{k=-N}^{N}\sum_{j=1}^4
  e_k^{(j)}\frac{\partial u_k^{(j)}}{\partial g^{(i_0)}_{m_0}[l_0]},
\end{equation}
where $l_0\in\{0, 1, \cdots,L_g-1\}$, $m_0\in\{0, 1, \cdots,M-1\}$,
and $i_0\in\{1,2,3,4\}$.  From the slicer error $e_{k}^{(j)}$
specified by~\eqref{eq:ePC}, define the $T_s=T/2$ oversampled slicer
error as
\begin{equation}
  e^{(j)}[n] = 
  \begin{cases} 
    e_{n/2}^{(j)}              & \mbox{if } n= 0,\pm 2,\pm 4,\cdots   \\
    0 & \mbox{otherwise}
  \end{cases}.
\end{equation}
Then, \eqref{eq:dEdg} can be rewritten as
\begin{equation}
  \label{eq:dEdg2}
  \frac{\partial { \overline {\mathcal E}_N}}{\partial g^{(i_0)}_{m_0}[l_0]}=\frac{2}{2N+1}\sum_{n=-2N}^{2N}\sum_{j=1}^4
  e^{(j)}[n]\frac{\partial u^{(j)}[n]}{\partial g^{(i_0)}_{m_0}[l_0]},
\end{equation}
where $u^{(j)}[n]$ is the oversampled output of the DSP block given by
\begin{equation}
  \label{eq:un}
  u^{(j)}[n]=\sum_{i=1}^4\sum_{l=0}^{L_{\Gamma}-1}
  {\Gamma}^{(j,i)}_{n}[l]{x}^{(i)}[n-l],\quad j=1,\cdots,4.
\end{equation}
The time index $n$ can be expressed as
\begin{equation}
  \label{eq:n}
  n=m+k'M,\quad m=0,1,\cdots,M-1;\quad \forall k',
\end{equation}
with $k'$ integer. Then, omitting the constant factor
$\frac{2}{2N+1}$, we can express the derivative~\eqref{eq:dEdg2} as
\begin{equation}
  \label{eq:dEdg3}
  \frac{\partial { \overline {\mathcal E}_N}}{\partial g^{(i_0)}_{m_0}[l_0]}\propto\sum_{k'}\sum_{m=0}^{M-1}\sum_{j=1}^4
  e^{(j)}[m+k'M]\frac{\partial u^{(j)}[m+k'M]}{\partial g^{(i_0)}_{m_0}[l_0]}.
\end{equation}

Next we evaluate the derivative
$\frac{\partial u^{(j)}[m+k'M]}{\partial g^{(i_0)}_{m_0}[l_0]}$.
Considering that the DSP filter coefficients $\Gamma^{(j,i)}_{n}[l]$
and the CE coefficients $g^{(i_0)}_{m_0}[l_0]$ are independent,
from~\eqref{eq:un} and~\eqref{eq:n} we verify that
\begin{equation}
  \label{eq:dundg}
  \frac{\partial u^{(j)}[m+k'M]}{\partial g^{(i_0)}_{m_0}[l_0]}=\sum_{i=1}^4\sum_{l=0}^{L_{\Gamma}-1}
  {\Gamma}^{(j,i)}_{m+k'M}[l]\frac{\partial {x}^{(i)}[m+k'M-l]}{\partial g^{(i_0)}_{m_0}[l_0]}.
\end{equation}

By using~\eqref{eq:n}, the signal at the $i$-th DSP block input given
by~\eqref{eq:eq6b} can be rewritten as
\begin{equation}
  x^{(i)}[m+k'M]=\sum_{l'=0}^{L_g-1} {g}^{(i)}_{m}[l']
  {w}^{(i)}[m+k'M-l'].
\end{equation}
Therefore,
\begin{equation}
  \label{eq:dxdg}
  \frac{\partial x^{(i)}[m+k'M]}{\partial g^{(i_0)}_{m_0}[l_0]}=
  {w}^{(i)}[m+k'M-l_0]\delta_{m,m_0}\delta_{i,i_0},
\end{equation}
where $\delta_{n,m}$ is the Kronecker delta function (i.e.,
$\delta_{n,m}=1$ if $n=m$ and $\delta_{n,m}=0$ if $n\ne m$).
Replacing \eqref{eq:dxdg} in \eqref{eq:dundg} we obtain
\begin{align}
  \nonumber
  &\frac{\partial u^{(j)}[m+k'M]}{\partial g^{(i_0)}_{m_0}[l_0]}=\\
  \label{eq:dudg2}
  &\quad\quad\quad \sum_{l=0}^{L_{\Gamma}-1} {\Gamma}^{(j,i_0)}_{m+k'M}[l] {w}^{(i_0)}[m+k'M-l-l_0]\delta_{m,m_0}.
\end{align}
Then introducing~\eqref{eq:dudg2} in~\eqref{eq:dEdg3}, we get
\begin{align}
  \frac{\partial { \overline {\mathcal E}_N}}{\partial g^{(i_0)}_{m_0}[l_0]}\propto&\sum_{k'}\sum_{j=1}^4
  e^{(j)}[m_0+k'M]\times\\
\nonumber
    &\sum_{l=0}^{L_{\Gamma}-1} {\Gamma}^{(j,i_0)}_{m_0+k'M}[l] {w}^{(i_0)}[m_0+k'M-l-l_0].
\end{align}
Finally, we set $kM=k'M-l$ resulting
\begin{equation}
  \label{eq:dEdg4}
  \frac{\partial { \overline {\mathcal E}_N}}{\partial g^{(i_0)}_{m_0}[l_0]}\propto\sum_{k}{\hat e}^{(i_0)}[m_0+kM] {w}^{(i_0)}[m_0+kM-l_0],
\end{equation}
where
\begin{equation}
  \label{eq:epb}
  {\hat e}^{(i)}[n]=\sum_{j=1}^4\sum_{l=0}^{L_{\Gamma}-1} {\Gamma}^{(j,i)}_{n+l}[l]e^{(j)}[n+l]
\end{equation}
is the backpropagated error. Notice that~\eqref{eq:dEdg4} is the
average of the instantaneous gradient component given by
${\hat e}^{(i_0)}[m_0+kM] {w}^{(i_0)}[m_0+kM-l_0]$.  As a consequence,
we can write an instantaneous gradient of the square error as
\begin{equation}
  \nabla_{{\mathbf g}^{(i)}_{m}} {\mathcal E}_k\propto{\hat e}^{(i)}[m+kM]{\mathbf w}^{(i)}[m+kM],
\end{equation}
with ${\mathbf w}[n]$ being the $L_g$-dimensional vector with the
samples at the CE input defined by~\eqref{eq:vecw}.

\section*{Acknowledgment}
The authors would like to thank MOSIS for fabricating their design
through the MEP research program.

\bibliographystyle{IEEEtran}
\bibliography{IEEEabrv,journal_bp_cal}


\end{document}